\documentclass{article}

\usepackage{microtype}
\usepackage{graphicx}
\usepackage{subfigure}
\usepackage{booktabs} 

\usepackage{hyperref}

\usepackage[accepted]{icml2024}

\usepackage{amsmath}
\usepackage{amssymb}
\usepackage{mathtools}
\usepackage{amsthm}
\usepackage{bbm}

\usepackage[capitalize,noabbrev]{cleveref}

\theoremstyle{plain}
\newtheorem{theorem}{Theorem}[section]

\newtheorem{lemma}[theorem]{Lemma}
\newtheorem{corollary}[theorem]{Corollary}
\theoremstyle{definition}

\newtheorem{assumption}[theorem]{Assumption}
\theoremstyle{remark}

\newcommand{\ceil}[1]{\left\lceil #1 \right\rceil}

\DeclareMathOperator*{\argmin}{arg\,min}

\newcommand{\indep}{\perp \!\!\! \perp}

\newlength\myindent
\usepackage[textsize=tiny]{todonotes}

\icmltitlerunning{Variational Inference with Coverage Guarantees in Simulation-Based Inference}

\begin{document}

\twocolumn[
\icmltitle{Variational Inference with Coverage Guarantees in Simulation-Based Inference}



\icmlsetsymbol{equal}{*}


\begin{icmlauthorlist}
\icmlauthor{Yash Patel}{umich}
\icmlauthor{Declan McNamara}{umich}
\icmlauthor{Jackson Loper}{umich}
\icmlauthor{Jeffrey Regier}{equal,umich}
\icmlauthor{Ambuj Tewari}{equal,umich}
\end{icmlauthorlist}

\icmlaffiliation{umich}{Department of Statistics, University of Michigan, Ann Arbor, USA}

\icmlcorrespondingauthor{Yash Patel}{yppatel@umich.edu}

\icmlkeywords{Variational Inference, Conformal Prediction, Calibration}

\vskip 0.3in
]



\printAffiliationsAndNotice{\icmlEqualContribution} 

\begin{abstract}
Amortized variational inference is an often employed framework in simulation-based inference that produces a posterior approximation that can be rapidly computed given any new observation. Unfortunately, there are few guarantees about the quality of these approximate posteriors. We propose Conformalized Amortized Neural Variational Inference (CANVI), a procedure that is scalable, easily implemented, and provides guaranteed marginal coverage. Given a collection of candidate amortized posterior approximators, \textsc{CANVI} constructs conformalized predictors based on each candidate, compares the predictors using a metric known as predictive efficiency, and returns the most efficient predictor. \textsc{CANVI} ensures that the resulting predictor constructs regions that contain the truth with a user-specified level of probability. \textsc{CANVI} is agnostic to design decisions in formulating the candidate approximators and only requires access to samples from the forward model, permitting its use in likelihood-free settings. We prove lower bounds on the predictive efficiency of the regions produced by \textsc{CANVI} and explore how the quality of a posterior approximation relates to the predictive efficiency of prediction regions based on that approximation. Finally, we demonstrate the accurate calibration and high predictive efficiency of \textsc{CANVI} on a suite of simulation-based inference benchmark tasks and an important scientific task: analyzing galaxy emission spectra.
\end{abstract}

\section{Introduction}

In many scientific applications, such as in astrophysics, neuroscience, and particle physics \citep{papamakarios2016fast,lueckmann2017flexible,greenberg2019automatic,deistler2022truncated,papamakarios2019sequential,boelts2022flexible}, posterior distributions $\mathcal{P}(\Theta\mid x)$ are sought over a large collection of $x$, typically on the order of 10,000 or more. In such scientific settings, $(\theta, x)$ pairs are assumed to come from nature, which has led to the growth of ``simulation-based inference,'' in which a likelihood $\mathcal{P}(X\mid\theta)$ and prior $\mathcal{P}(\Theta)$ are posited and simulated to fit posteriors \citep{cranmer2020frontier,lueckmann2021benchmarking}. Even when the likelihood and prior are well specified, exact sampling from the posteriors is intractable, requiring running 10,000 separate MCMC chains.

In these settings, amortized variational inference is frequently employed. Variational inference (VI) has become a staple in Bayesian inference; however, it has been repeatedly noted that a major shortcoming of VI is its lack of any theoretical guarantees and tendency to produce biased posterior estimates \citep{blei2017variational, murphy2022probabilistic, zhang2018advances, yao2018yes}. 
The most common metric for assessing the calibration of such inference algorithms is the ``expected coverage.'' Having calibrated expected coverage is a necessary but not sufficient condition for conditional coverage, yet amortized variational approximations fail to even achieve this minimal requirement, despite significant work to remedy this shortcoming \citep{deistler2022truncated,delaunoy2022towards,lemos2023sampling,delaunoy2023balancing}. A lack of such calibration limits the capacity to reach downstream scientific conclusions, highlighted in a recent meta-study of likelihood-free inference algorithms \citep{hermans2021averting}.

\begin{figure*}
\centering
\includegraphics[scale=0.20]{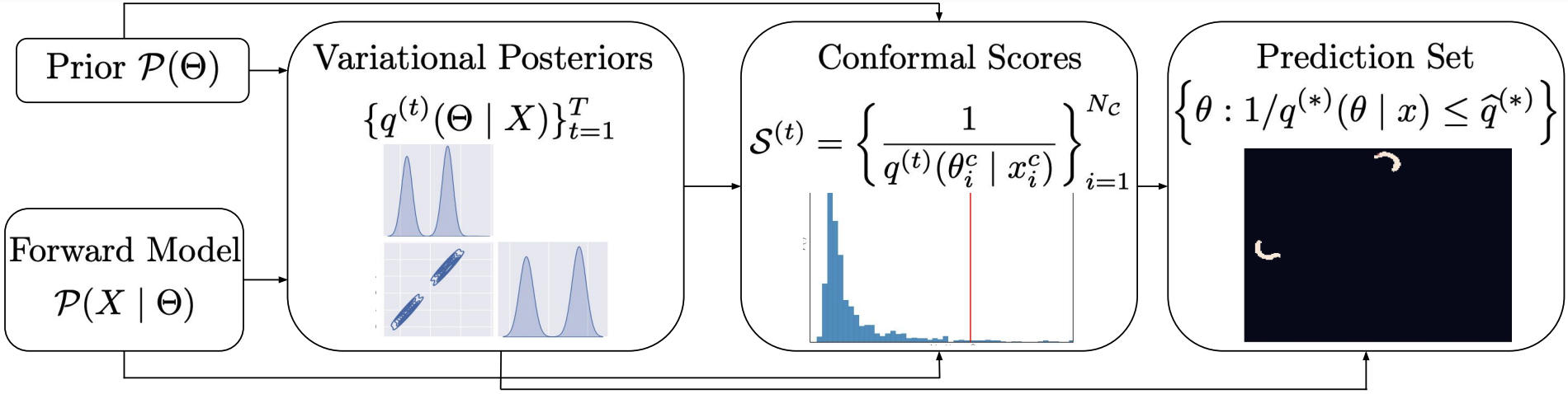}
\caption{\label{fig:workflow} \textsc{CANVI} is a wrapper around variational inference requiring minimal implementation and computational overhead that produces prediction regions with guaranteed marginal calibration. Among a family of candidate amortized posterior approximators, \textsc{CANVI} can identify the approximator leading to the most efficient prediction regions. \textsc{CANVI} can be used in any setting where the forward model $\mathcal{P}(X \mid \Theta)$ can be sampled.}
\end{figure*}

In applications where many posteriors need to be estimated, credible regions with \textit{marginally} calibrated coverage can be sufficient for downstream scientific inquiries. For instance, in astrophysics, there is great interest in constraining the $\Lambda$CDM model, the current concordance model in cosmology \citep{gilman2021strong, hezaveh2016detection, vegetti2010detection, vegetti2009statistics, hogg1994gravitational}. A recent work from this community, \citep{hermans2021towards}, leveraged Bayesian inference towards this end, obtaining approximate posteriors for the parameters of interest on each of 10,000 observations. Crucially, these posteriors were then used to produce credible intervals with \textit{marginally} valid frequentist coverage, from which they made claims on the $\Lambda$CDM model.



Our insight is that conformal prediction can be leveraged to provide variational approximators with marginal coverage guarantees and provide users new ways to measure the quality of variational approximators. In this manuscript, we present \textsc{CANVI} (Conformalized Amortized Neural Variational Inference), a novel, general framework for producing marginally calibrated, informative prediction regions from a collection of variational approximators. Such regions can be produced with minimal implementation and computational overhead, requiring only samples from the prior and $\mathcal{P}(X \mid \Theta)$, as shown in Figure \ref{fig:workflow}. In Section \ref{section:canvi_eff}, we provide theoretical analysis of the informativeness of the prediction regions produced by \textsc{CANVI} using a measure known as ``predictive efficiency.'' High predictive efficiency is necessary to draw conclusions in downstream scientific inquiries and relates to asymptotic conditional coverage with the appropriate choice of score function. Finally, in Section \ref{section:experiments}, we show calibration and predictive efficiency across simulation-based inference benchmark tasks and an important scientific task: analyzing galaxy emission spectra.

\section{Background}\label{section:background}
\subsection{Variational Inference}
Bayesian methods aim to sample the posterior distribution $\mathcal{P}(\Theta \mid X)$, typically using either MCMC or VI. VI has risen in popularity recently due to how well it lends itself to amortization. Given an observation $X$, variational inference transforms the problem of posterior inference into an optimization problem by seeking
\begin{equation}
\label{eq:whatisvi}
\varphi^*(X) = \argmin_{\varphi} D(q_{\varphi}(\Theta) || \mathcal{P}(\Theta\mid X)),
\end{equation}
where $D$ is a divergence and $q_{\varphi}$ is a member of  a variational family of distributions $\mathcal{Q}$ indexed by the free parameter $\varphi$. Normalizing flows have emerged as a particularly apt choice for $\mathcal{Q}$, as they are highly flexible and perform well empirically \citep{rezende2015variational, agrawal2020advances}. Amortized variational inference expands on this approach by training a neural network to approximate $\varphi^*(X)$. This leads to a variational posterior approximator $q(\Theta\mid X)=q_{\varphi^*(X)}(\Theta)$ that can be rapidly computed for any value $X$.   
The characteristics of $\varphi^*$ depend in part on the variational objective, $D$. 
For instance, using a reverse-KL objective, i.e. $D_{KL}(q_{\varphi}(\Theta) || \mathcal{P}(\Theta\mid X))$, is known to produce mode-seeking posterior approximations, whereas using a forward-KL objective, i.e. $D_{KL}(\mathcal{P}(\Theta\mid X)||q_{\varphi}(\Theta))$, encourages mode-covering behavior \citep{pml2Book}. Popular variational objectives include the Forward-Amortized Variational Inference (FAVI) objective \citep{ambrogioni2019forward, bornschein2014reweighted}, the Evidence Lower Bound (ELBO), and the Importance Weighted ELBO (IWBO) \citep{burda2015importance}.

\subsection{Conformal Prediction}
Given $\mathcal{D_C}=\{(x_1,\theta_1),\ldots (x_{N},\theta_{N})\}\stackrel{\text{iid}}{\sim}\mathcal{P}(X, \Theta)$, 
conformal prediction \citep{angelopoulos2021gentle, shafer2008tutorial} produces prediction regions with distribution-free guarantees. A prediction region is a mapping from observations of $X$ to sets of possible values for $\Theta$ and is said to be marginally calibrated at the $1-\alpha$ level if $\mathcal{P}(\Theta \notin \mathcal{C}(X))\leq \alpha$.

Split conformal is one popular version of conformal prediction. In this approach, marginally calibrated regions $\mathcal{C}$ are designed using a ``score function'' $s(x,\theta)$. Intuitively, the score function should have the quality that $s(x,\theta)$ is smaller when it is more reasonable to guess that $\Theta=\theta$ given the observation $X=x$. For example, if one has access to a function $\hat f(x)$ which attempts to predict $\Theta$ from $X$, one might take $s(x,\theta)=\Vert\hat f(x)-\theta\Vert$. The score function is evaluated on each point of a subset of the dataset $\mathcal{D_C}$, called the ``calibration dataset,'' yielding $\mathcal{S} = \{s(x_i^c, \theta_i^c)\}_{i=1}^{N_{\mathcal{C}}}$. Note that the calibration dataset cannot be used to pick the score function; if data is used to design the score function, it must independent of  $\mathcal{D_C}$. This is how ``split conformal'' gets its name: in typical cases, data are split into two parts, one used to design $s$ and the other to perform calibration. We then define $\widehat q(\alpha)$ as the $\ceil{(N_{\mathcal{C}}+1)(1-\alpha)}/N_{\mathcal{C}}$ quantile of $\mathcal{S}$. For any future $x$, the set $\mathcal{C}(x) = \{\theta \mid s(x, \theta) \le \widehat{q}(\alpha)\}$ satisfies $1 - \alpha\le\mathcal{P}(\Theta \in \mathcal{C}(X))$. This coverage guarantee arises from the exchangeability of the score of a future test point $s(x', \theta')$ with $\mathcal{S}$ and holds regardless $N_{\mathcal{C}}$: conformal guarantees are not asymptotic. If $s(X,\Theta)$ is jointly continuous, we additionally have that $\mathcal{P}(\Theta \in \mathcal{C}(X))\le 1 - \alpha + 1/(N_{\mathcal{C}} + 1)$.

While the coverage guarantee holds for any score function, different choices may lead to more or less informative prediction regions \citep{shafer2008tutorial}. For example, the score $s(x,\theta)=1$ leads to the uninformative prediction region of all possible values of $\Theta$. Predictive efficiency is one way to quantify informativeness \citep{yang2021finite,sesia2020comparison}, defined as the expected inverse Lebesgue measure of $\mathcal{C}(X)$, $\mathbb{E}_{X}[\mathcal{L}(\mathcal{C}(X))^{-1}]$. Methods using conformal prediction often seek to identify prediction regions that are efficient and marginally calibrated.

\subsection{Related Literature}\label{section:related}
Efforts to correct for miscalibration have been of great interest recently in light of its apparent omnipresence highlighted in \citep{hermans2021averting}. The notion of calibration studied therein, and the one we concentrate on here, centers on the expected coverage probability of the highest predictive density (HPD) of a posterior estimate $q(\Theta\mid X)$, defined for such a $q$, some fixed $x$, and pre-specified coverage level $\alpha$ to be the set $\mathrm{HPD}(q(\Theta\mid x),1-\alpha)$ with smallest Lebesgue measure such that $\mathcal{P}_{\Theta\sim q(\Theta\mid x)}(\Theta\in\mathrm{HPD}(q(\Theta\mid x),1-\alpha))\ge 1-\alpha$. The expected coverage is then $\mathbb{E}_{X,\Theta\sim\mathcal{P}(X,\Theta)}[\mathbbm{1}[\Theta\in\mathrm{HPD}(q(\Theta\mid X),1-\alpha)]]$.

Such calibration has been studied across a number of posterior estimation techniques in the likelihood-free inference community, which generally fall into one of three categories. Neural posterior approximation (NPE) methods directly approximate the posterior density $\mathcal{P}(\Theta\mid X)$ \citep{papamakarios2016fast,lueckmann2017flexible,greenberg2019automatic,deistler2022truncated}, neural likelihood estimation (NLE) methods estimate the assumed intractable likelihood $\mathcal{P}(X\mid \Theta)$ \citep{papamakarios2019sequential,boelts2022flexible}, and neural ratio estimation (NRE) methods estimate the $\mathcal{P}(X\mid\Theta) / \mathcal{P}(X)$ ratio \citep{hermans2020likelihood,durkan2020contrastive,miller2022contrastive}. Both NLE and NRE rely on MCMC for posterior sampling after estimation.

While these approaches differ in their estimation strategies, they all suffer miscalibration if naively employed. One suggestion advocated by \citep{hermans2021averting} was ensembling, compatible with all the aforementioned posterior approximation strategies. Despite its improvement in empirical calibration, ensembling affords no guarantees on calibration and also dramatically increases the computational cost of training and inference alike. As an effort to address this lack of guarantees and computational cost, recent calibration efforts have focused on modifying the loss function to appropriately encourage conservatism in the learned posterior estimates, since the downstream scientific use cases can afford such conservatism, unlike underdispersed posteriors.
In particular, \citep{delaunoy2022towards} took a first step in this direction by proposing a modification over the vanilla NRE formulation with a regularization term that results in more conservative posterior estimates. Such a method, however, has no guarantees and further
relies on the selection of a tunable $\lambda$ parameter, whose optimal selection is unknown without awareness of the true posterior. This approach was then extended to NPE and NLE in \citep{delaunoy2023balancing} and \citep{falkiewicz2023calibrating}, which both again suffer from the deficiencies of producing overly conservative posteriors and relying on the careful selection of $\lambda$.


\section{Method}\label{section:method}
In response to the shortcomings highlighted in the previous section, we were interested in procuring a method that has (1) guarantees on calibration without tuning parameters, (2) minimal computational overhead, and (3) informative prediction regions. We demonstrate in \Cref{section:canvi_framework} that leveraging conformal prediction on an amortized VI approximator $q(\Theta\mid X)$ immediately addresses points (1) and (2).  
We then study in the following sections how this naive application can be extended to the full \textsc{CANVI} algorithm by considering a collection of amortized VI approximators $\{q^{(1)}(\Theta\mid X),\ldots, q^{(T)}(\Theta\mid X)\}$ to address point (3).
CANVI can be applied whenever $\mathcal{P}(X, \Theta)$ can be sampled. The coverage validity of \textsc{CANVI} is proven in Section \ref{section:canvi_approx_selection}, and analyses of its predictive efficiency in Section \ref{section:efficiency_proof}. 

\subsection{CANVI: Score Function}\label{section:canvi_framework}
In the simplest case, \textsc{CANVI} takes as input a single amortized posterior approximator $q(\Theta\mid X)$.
In traditional applications of split conformal, much concern is given to the loss of accuracy of the predictor $q$ in having to reserve a subset of the training data for calibration.
Here we have no such issues; we sample $\mathcal{D_C} = \{(x_i^c, \theta_i^c)\}_{i=1}^{N_{\mathcal{C}}}\overset{\mathrm{iid}}{\sim}\mathcal{P}(\Theta)\mathcal{P}(X\mid \Theta)$ from the joint distribution to produce a calibration dataset that can be arbitrarily large. Given $q(\Theta\mid X)$, we employ the following score, as used in \citep{angelopoulos2021gentle}:
\begin{equation}\label{eqn:canvi_score}
    s(x_i,\theta_i) = (q(\theta_i\mid x_i))^{-1}.
\end{equation}
Denoting the $\ceil{(N_{\mathcal{C}}+1)(1-\alpha)}/N_{\mathcal{C}}$ quantile of the score distribution over $\mathcal{D}_{C}$ as $\widehat{q}_\mathcal{C}(\alpha)$, $\mathcal{C}(x)=\{\theta : 1 / q(\theta\mid x) \le \widehat{q}_\mathcal{C}(\alpha)\}$ is then marginally calibrated. It may be disjoint if the posterior is multimodal.

While other choices of score functions also result in regions $\mathcal{C}(x)$ that address both the lack of guarantees and computational cost of methods highlighted in \Cref{section:related}, the particular choice of \Cref{eqn:canvi_score} has the desirable property that, if we recover the true posterior, that is $q(\Theta\mid X) = \mathcal{P}(\Theta\mid X)$, we recover the HPDs, namely $\mathcal{C}(x) = \mathrm{HPD}(\mathcal{P}(\Theta\mid x),1-\alpha)$, achieving conditional coverage. 
Note that the procedure described in this section and those that follow can be easily extended to the group conditional setting if such coverage is of interest, whose discussion we defer to \Cref{section:group_cond}.

\subsection{CANVI: Approximator Selection}\label{section:canvi_approx_selection}
To mitigate the risk of producing uninformative prediction regions, it is natural to explore multiple posterior approximations, $\{q^{(t)}(\Theta\mid X)\}_{t=1}^T$, since a poorly chosen approximator may lead to poor predictive efficiency. These posterior approximations could, for instance, differ in their choice of training objective, variational family, or hyperparameters. \textsc{CANVI} seeks to identify the variational approximator $q^{(t^{*})}$ for which the efficiency is greatest, defined for a threshold $\tau$ to be
\begin{equation}\label{eqn:cp_global_eff}
    \ell(q, \tau) := \mathbb{E}_X\left[\mathcal{L}\left(\left\{\theta : 1/ q(\theta\mid X) \le \tau\right\}\right)\right],
\end{equation}
We defer the discussion of estimating $\ell(q, \tau)$ to Section \ref{section:canvi_vol_est}. Naively, one would expect by taking $(q^{(t^{*})}, \widehat{q}_\mathcal{C}^{(t^{*})}(\alpha))$ where $t^{*} := \argmin_{t} \ell(q^{(t)}, \widehat{q}_\mathcal{C}^{(t)}(\alpha))$, we achieve maximal efficiency and retain coverage guarantees. However, defining $\mathcal{C}(x)$ with $\widehat{q}_\mathcal{C}^{(t^{*})}(\alpha)$ fails to retain coverage guarantees, as the exchangeability of scores of future test points $s(x', \theta')$ with $\mathcal{S}$ is lost in conditioning on $\mathcal{D_C}$ for selecting $t^{*}$. 

We must, therefore, perform an \textit{additional} recalibration step after selecting $t^{*}$ to retain coverage guarantees. \textsc{CANVI} performs such recalibration using an additional dataset $\mathcal{D_R}$ again constructed with i.i.d. draws from $\mathcal{P}(X, \Theta)$. We take $\mathcal{D_R}$ to be the same size as $\mathcal{D_C}$, i.e. $|\mathcal{D_R}| = N_{\mathcal{C}}$. \textsc{CANVI} then computes the quantile $\widehat{q}_{\mathcal{R}}^{(*)}(\alpha) := \widehat q_\mathcal{R}^{(t^{*})}(\alpha)$, which is used to define prediction regions $\mathcal{C}(x) := \{\theta : 1/ q^{(t^{*})}(\theta\mid x) \le \widehat{q}_{\mathcal{R}}^{(*)}(\alpha)\}$. However, such regions require analysis to guarantee high efficiency, as we explore in Section \ref{section:canvi_eff}.



The full \textsc{CANVI} framework is provided in Algorithm \ref{alg:canvi}.
The validity of the \textsc{CANVI} procedure follows directly from that of split conformal prediction, formally stated below and explicitly proven in Appendix \ref{section:canvi_validity}.

\begin{lemma}
    Let $\alpha\in(0,1)$ and
    \begin{gather*}
        q^{(*)}(\Theta\mid X),\widehat{q}_{\mathcal{R}}^{(*)}(\alpha) = \\
        \mathrm{CANVI}\left(
        \{q^{(t)}(\Theta\mid X)\}_{t=1}^T,\mathcal{P}(X, \Theta),1-\alpha,N_{\mathcal{C}},N_{\mathcal{T}}
    \right)
    \end{gather*}
    Let $(x', \theta')\sim\mathcal{P}(X, \Theta) \indep \mathcal{D} \cup \mathcal{D_C} \cup \mathcal{D_R} \cup \mathcal{D_T}$, with $\mathcal{D}$ being the data used to train $\{q^{(t)}(\Theta|X)\}_{t=1}^T$. Then $1 - \alpha\le\mathcal{P}(1/q^{(*)}(\theta'|x') \leq \widehat{q}_{\mathcal{R}}^{(*)}(\alpha))$.
\end{lemma}

\begin{algorithm}
   \caption{\textsc{CANVI}: Note that \textsc{VolumeEst} is a volume estimator subroutine detailed in Section \ref{section:canvi_vol_est}.}
   \label{alg:canvi}
    \begin{algorithmic}[1]
    \STATE{\textbf{Procedure} CPQuantile}{}
    \STATE{\textbf{Inputs: } Posterior approximation $q(\Theta\mid X)$, Calibration set $\mathcal{D_C}$, Desired coverage $1-\alpha$}
    \STATE{$\mathcal{S} \gets \{\frac{1}{q(\theta_i\mid x_i)}\}_{i=1}^{N_{\mathcal{C}}}$}
    \STATE{\textbf{Return} $\frac{\ceil{(N_{\mathcal{C}}+1)(1-\alpha)}}{N_{\mathcal{C}}} \text{ quantile of } \mathcal{S}$}
    \STATE 
    \STATE{\textbf{Procedure} \textsc{CANVI}}{}
    \STATE \textbf{Inputs: } Posterior approximators $\{q^{(t)}(\Theta|X)\}_{t=1}^T$, Prior $\mathcal{P}(\Theta)$, Forward model $\mathcal{P}(X\mid \Theta)$,  Desired coverage $1-\alpha$, Calibration size $N_{\mathcal{C}}$, Test size $N_{\mathcal{T}}$
    \STATE{$\mathcal{D_C},\mathcal{D_R} \sim \mathcal{P}(X, \Theta), \mathcal{D_T}\sim\mathcal{P}(X)$} 
    \FOR{$t \in\{1,\ldots T\}$}
    \STATE{$\widehat{q}_{\mathcal{C}}^{(t)} \gets \textsc{CPQuantile}(q^{(t)}, \mathcal{D_C}, 1-\alpha)$}
    \STATE{$\widehat{\ell}^{(t)} \gets \textsc{VolumeEst}(q^{(t)}, \mathcal{P}(\Theta), \widehat{q}_{\mathcal{C}}^{(t)}, \mathcal{D_T})$}
    \ENDFOR
    \STATE{$t^{*} \gets \argmin_t \widehat{\ell}^{(t)}$} 
    \STATE{$\widehat{q}_{\mathcal{R}}^{(*)}(\alpha) \gets \textsc{CPQuantile}(q^{(t^{*})}, \mathcal{D_R}, 1-\alpha)$}
    \STATE{\textbf{Return} $q^{(*)}(\Theta\mid X)$, $\widehat{q}_{\mathcal{R}}^{(*)}(\alpha)$} 
    \end{algorithmic}
\end{algorithm}



\subsection{CANVI: Efficiency Analysis Assumptions}\label{section:canvi_eff}
We now show that, with high probability, the pair \textsc{CANVI} produces $(q^{(*)}(\Theta| X),\widehat{q}_{\mathcal{R}}^{(*)}(\alpha))$ is the most efficient amongst the candidate posteriors considered. The concern is that the post-recalibrated quantile may result in significant degradation of the efficiency, i.e. $\ell(q^{(*)}, \widehat{q}_{\mathcal{R}}^{(*)}(\alpha)) \gg \ell(q^{(*)}, \widehat{q}_{\mathcal{C}}^{(*)}(\alpha))$. This tradeoff between coverage and efficiency was studied in \citep{yang2021finite}.

Recall $1 - \alpha\le\mathcal{P}(\Theta \in \mathcal{C}(X))\le 1 - \alpha + 1/(N_{\mathcal{C}} + 1)$. Denote the CDF of the score function under the joint distribution $\mathcal{P}(X, \Theta)$ as $\mathcal{F}(s) := \mathcal{P}_{\Theta,X}(1/q(\Theta\mid X))$. The coverage guarantee, thus, implies 
$\widehat{q}(\alpha) \in [\mathcal{F}^{-1}(1 - \alpha), \mathcal{F}^{-1}(1 - \alpha + 1/(N_{\mathcal{C}} + 1))]$ for $\widehat{q}(\alpha)$ from \textit{any} calibration set. In particular, $\widehat{q}_{\mathcal{C}}^{(*)}(\alpha)$ and $\widehat{q}_{\mathcal{R}}^{(*)}(\alpha)$ both lie in this range.

We bound the efficiency suboptimality by proceeding in two steps. We first demonstrate that the quantiles of $q^{(*)}(\Theta\mid X)$ under $\mathcal{D}_\mathcal{C}$ and $\mathcal{D}_\mathcal{R}$ are close by demonstrating the quantile range of $\mathcal{F}^{-1}$ is small. We then demonstrate the efficiency varies smoothly as a function of the quantile, allowing us to bound the resulting efficiency change. Formally, we state these assumptions respectively as follows, per \citep{yang2021finite}, which we then demonstrate follow from properties of the chosen variational families.


\begin{assumption}\label{assump:opt_technical_2}
    For each $t$, the $\ell(q^{(t)}, \tau)$ is Lipschitz continuous in $\tau$ with constant $L_W$.
\end{assumption}

\begin{assumption}\label{assump:opt_technical_1}
    For each $t$, $\exists$ $r, \gamma\in(0,1]$ such that $\mathcal{F}_t^{-1}(s)$ (inverse score CDF under $q^{(t)}$), is $\gamma$-Hölder continuous on \upshape[$ 1-\alpha,1-\alpha + r$\upshape] \itshape with continuity constant $L_t$.
\end{assumption}



\subsubsection{Lipschitz Continuity of Efficiency}
We now demonstrate Assumption \ref{assump:opt_technical_2}
can be guaranteed with the appropriate selection of variational family by the end user. It suffices to demonstrate $\ell_x(q, \tau) := \mathcal{L}(\{\theta : 1/ q(\theta\mid x) \le \tau\})$ is $L$-Lipschitz continuous in $\tau$ for any $x\in \mathcal{X}$, proven in Appendix \ref{section:lip_cts_expectation}. 

For any fixed $x$, $\ell_x(q, \tau)$ is the intrinsic volume of the sublevel set of $1 / q(\theta\mid x)$. 
We summarize and subsequently use relevant results from \citep{jubin2019intrinsic} below. ``Intrinsic volume'' defines the notion of volume for a lower dimensional manifold embedded in a higher dimensional space. For a brief review of Riemannian manifolds, see Appendix \ref{section:riemannian_review}; a more complete presentation is available in \citep{lee2012smooth}. The $n-k$ degree intrinsic volume of a flat compact $n$-dimensional manifold $N$ is
\begin{equation}
    \mathcal{L}_{n-k}(N) = b_{k} \int_{\partial N} \text{tr}\left(\bigwedge\nolimits^{k-1} S\right) \text{vol}_{\partial N},
\end{equation}
where $0\le k\le n$, $b_{k}\in\mathbb{R}$, and $S$ is the second fundamental form of $\partial N$ in $N$. Lipschitz continuity of $\mathcal{L}_{n-k}(M_{f}^{\tau})$ was established as follows. The level sets $M_{f}^{\tau} := f^{-1}((-\infty, \tau])$ are restricted to ``regular values'' of $f$, namely $\tau$ such that $f(x) = \tau \implies df(x)\neq 0$.

\begin{theorem}\label{thm:lip_cts}
    Let $(M, g)$ be an $n$-dimensional Riemannian manifold, $f\in\mathcal{C}^{3}(M, \mathbb{R})$ bounded below, and $\tau$ a regular value of $f$ equipped with the standard uniform $\mathcal{C}^{3}$ topology. Then, if $0\le k\le n$, $\tau\rightarrow \mathcal{L}_{n-k}(M_{f}^{\tau})$ is Lipschitz continuous \citep{jubin2019intrinsic}.
\end{theorem}

The Lipschitz continuity of $\ell_{x}(q, \tau)$ then follows as a corollary for any variational families for which the density is sufficiently smooth, namely $q(\theta\mid x)\in\mathcal{C}^{3}(\mathbb{R}^{n})$. The proof follows by taking $\ell_{x}(q, \tau) := \mathcal{L}_{n}((\mathbb{R}^{n})_{s}^{\tau})$ for $s(\theta) = 1 / q(\theta|x)$, with $\tau=s(\theta)$ for $\{\theta : \nabla_{\theta} q(\theta|x) \neq 0\}$ being the set of regular values. This domain restriction is discussed more in Section \ref{section:efficiency_proof}. Notice $s(\theta)\in\mathcal{C}^{3}(\mathbb{R}^{n})$ as both $f(x) := 1 / x$ and $q(\theta\mid x)$ are $\in\mathcal{C}^{3}(\mathbb{R}^{n})$ and $\mathcal{C}^{3}$ is closed under function composition. Formally,

\begin{corollary}\label{corollary:smoothness}
    Suppose for any $x\in\mathcal{X}$, $q(\theta\mid x)\in\mathcal{C}^{3}(\mathbb{R}^{n})$ is bounded above and $\tau$ is a regular value of $q(\theta\mid x)$. Then, $\ell(q, \tau)$ is Lipschitz continuous in $\tau$.
\end{corollary}

Notably, this assumption on smoothness holds for most variational families used in practice, including highly expressive flow-based variational families \citep{kohler2021smooth}.

\subsubsection{Continuity of Conformal Quantiles}
We now discuss the validity of Assumption \ref{assump:opt_technical_1}. Comparable assumptions are commonly used in the quantile estimation literature, as discussed in \citep{lei2018distribution}. 
The Hölder constant cannot be characterized in general, as it is intimately tied to specific details of the score distribution under $\mathcal{P}(X, \Theta)$. We, therefore, provide an explicit characterization for a particular family of distributions in Theorem \ref{thm:gaussian_cont} and defer extensions to a broader set of families to future work. Details for this proof are given in Appendix \ref{section:informative_thm_detailed}.



    


\begin{theorem}\label{thm:gaussian_cont}
Let $\Theta$ and $X$ be zero-mean unit-variance Gaussian random variables with correlation $\rho$. Let $q^{(t)}(\theta|x)=\mathcal{N}(\theta;t x, 1 - \rho^2)$. Let $\kappa := t^2 - 2 t \rho + 1$ and $r > 0$. Then $F_t^{-1}(z)$, is $1$-Hölder continuous on \upshape[$ 1-\alpha,1-\alpha + r$\upshape] \itshape with Hölder constant
\begin{equation}\label{eqn:gaussian_holder}
    \frac{\kappa \Phi^{-1}(\frac{1-\alpha}{2}) \sqrt{
    \exp\left(\frac{\kappa}{1-\rho^2} \Phi^{-1}(\frac{1-\alpha}{2})^2 -\frac{(1-\alpha)^2}{2}\right)
    }
    }{\sqrt{(1-\rho^2) / 2}}
\end{equation}
\end{theorem}


Notably, the Hölder constant is minimized in recovering the true posterior, as \Cref{eqn:gaussian_holder} is minimized at $\varphi = \rho$.




\subsection{CANVI: Volume Estimation}\label{section:canvi_vol_est}
We now provide an estimation procedure for $\ell(q, \tau)$. Naively, we might expect taking the sample average of $\ell_{x_{i}}(q, \tau)$ over $\mathcal{D_T} := \{x_i\}_{i=1}^{N_{\mathcal{T}}}\sim\mathcal{P}(X)$ would suffice. However, exact calculation of $\ell_{x_{i}}(q, \tau)$ requires a grid-discretization over $\text{Supp}(\Theta|x_i)$, which is only feasible when the support has a known, small extent. 



As a result, $\ell_{x}(q, \tau)$ is estimated using an importance-weighted Monte Carlo estimate over $S$ samples from $q$. Such an estimator, however, suffers from high variance if $q$ is underdispersed, as $\mathcal{C}(x)$ will cover regions of low variational density. To combat this issue, we use the well-known fact that $\mathcal{P}(\Theta \mid X)$ is narrower than $\mathcal{P}(\Theta)$ to construct a ``mixed sampler,'' specifically with $z_j\sim\text{Bern}(\lambda)$ and $\theta_j\sim q(\Theta|x_i)^{z_{j}} \mathcal{P}(\Theta)^{1-z_{j}}$, where the mixed density is now $\widetilde{q}(\theta_j) = \lambda q(\theta_{j}|x_i) + (1-\lambda)\mathcal{P}(\theta_{j})$. The necessity of such mixing is dependent on the nature of the variational posterior with respect to the true posterior, which is unknown in practice. We, thus, average several estimates over $\{\lambda_k\}\in[0,1]$. Denoting the mixed density with $\lambda_k$ as $\widetilde{q}_k$, for each $x_i$ and $\lambda_k$, we make $S$ draws $\{\theta_{jk}\}_{j=1}^{S}\sim \widetilde{q}_k(\Theta\mid x_i)$. We empirically demonstrate the necessity of such mixed sampling in Section \ref{section:training_obj}. This procedure is summarized in Algorithm \ref{alg:cp_region_size}, with the final estimate $\widehat{\ell}(q,\tau)$ being
\begin{equation}\label{eqn:cp_mc_est}
\frac{1}{K \mathcal{N_T}} \sum_{i,k=1}^{\mathcal{N_T},K}
\frac{1}{S} \sum_{j=1}^S \frac{1}{\widetilde{q}_k(\theta_{jk}|x_i)} \mathbbm{1}\left[ \frac{1}{q(\theta_{jk} | x_i)} \le \tau\right] 
\end{equation}




\begin{algorithm}
  \caption{\label{alg:cp_region_size} \textsc{VolumeEst}}
  \begin{algorithmic}[1]
    \STATE{\textbf{Procedure} VolumeEst}{}
    \STATE{\textbf{Inputs: } Posterior approximation $q(\Theta\mid X)$, Prior $\mathcal{P}(\Theta)$, CP quantile $\widehat{q}$, Test set $\mathcal{D_T}$}
    \FOR{$i \in\{1,\ldots \mathcal{N_T}\},k \in\{1,\ldots K\}$}
    \FOR{$j \in\{1,\ldots S\}$}
    \STATE{$z_{j} \sim\text{Bern}(k/K)$}
    \STATE{$\theta_{j}\sim q(\Theta \mid x_i)^{z_j} \mathcal{P}(\Theta)^{1-z_j}$}
    \STATE{$\widetilde{q}_j \gets (k/K) q(\theta_{j} \mid x_i) + (1-k/K)\mathcal{P}(\theta_{j})$}
    \ENDFOR
    \STATE{$V_{i,k} \gets \frac{1}{S} \sum_{j=1}^S \frac{1}{\widetilde{q}_j} \mathbbm{1}[ 1 / q(\theta_{j} | x_i) \le \widehat{q}]$}
    \ENDFOR
    \STATE{\textbf{Return} $\frac{1}{K \mathcal{N_T}} \sum_{i,k} V_{i,k}$}
    \end{algorithmic}
\end{algorithm}

\subsection{CANVI: Efficiency Proof}\label{section:efficiency_proof}

We now state the result of recovery of the optimal recalibrated approximator. To do so, we require the Monte Carlo estimate to be sufficiently well-behaved to recover the optimal pre-recalibration approximator.
\begin{assumption}\label{assump:opt_technical_3}
    If $t^{*} := \argmin_{1\le t\le T} \ell(q^{(t)}, \widehat{q}_\mathcal{R}^{(t)}(\alpha))$ and $\widehat{t}^{*} := \argmin_{1\le t\le T} \widehat{\ell}(q^{(t)}, \widehat{q}_\mathcal{R}^{(t)}(\alpha))$ for $\alpha\in(0,1)$, then $\exists \Delta, \epsilon > 0$, such that with probability at least $1-\epsilon$,
    $    |\ell(q^{(\widehat{t}^{*})}, \widehat{q}_\mathcal{R}^{(\widehat{t}^{*})}(\alpha)) - \ell(q^{(t^{*})}, \widehat{q}_\mathcal{R}^{(t^{*})}(\alpha))| < \Delta.
    $
\end{assumption}
Important to note is that $\widehat{\ell}$ is only used to select $t^{*}$, after which we make claims on $\ell$ (i.e. \textit{not} the estimate) for the recalibrated quantiles in Theorem \ref{thm:efficient}. As with Assumption \ref{assump:opt_technical_1}, this assumption is intimately tied to specific details of the score distribution under $\mathcal{P}(X, \Theta)$, making its restatement in more natural distributional properties of $q$ impossible. We demonstrate its validity empirically across several posteriors in Section \ref{section:empirical_eff}. 

The proof of Theorem \ref{thm:efficient} now follows as an extension of Theorem 3 from \citep{yang2021finite} and is explicitly provided in Appendix \ref{section:empirical_iterate_thm}. Notably, we use Corollary \ref{corollary:smoothness} to replace Assumption \ref{assump:opt_technical_2} with a more natural set of conditions for this context. This requires ensuring that, for any $x$, $\widehat{q}_{\mathcal{C}}^{(*)}(\alpha)$ and $\widehat{q}_{\mathcal{R}}^{(*)}(\alpha)$ are regular values of $s(\theta)$. We, thus, assume for any $x$ and $\theta\neq0$, $\mathcal{L}(\{\theta : \nabla_{\theta} q(\theta|x)\}) = 0$, which naturally holds for variational families used in practice, i.e. any non-piecewise constant density estimator.

\begin{theorem}\label{thm:efficient}
    Suppose for any $x\in\mathcal{X}$ and $t=1,...,T$, $q^{(t)}(\theta\mid x)\in\mathcal{C}^{3}(\mathbb{R}^{n})$ is bounded above and for $\theta\neq0$, $\mathcal{L}(\{\theta : \nabla_{\theta} q^{(t)}(\theta|x)\}) = 0$. Further assume $P(X,\Theta)$ is bounded above.
    Let $\alpha\in(0,1)$ and
    \begin{gather*}
        q^{(*)}(\Theta\mid X),\widehat{q}_{\mathcal{R}}^{(*)}(\alpha) = \\
        \mathrm{CANVI}\left(
        \{q^{(t)}(\Theta\mid X)\}_{t=1}^T,\mathcal{P}(X, \Theta),1-\alpha,N_{\mathcal{C}},N_{\mathcal{T}}
    \right)
    \end{gather*}    
    If, for
        $r\ge\max\{\sqrt{\log(4T/\delta)/2N_{\mathcal{C}}}, 2 / N_{\mathcal{C}}\}$
    and $\delta\in[0,1]$, Assumption \ref{assump:opt_technical_1} holds
    and for $\Delta, \epsilon > 0$ Assumption \ref{assump:opt_technical_3} holds, then with probability at least $(1-\epsilon)(1-\delta)$,
    \begin{equation}\label{eqn:iterate_thm}
    \begin{gathered}
        \ell(q^{(*)},\widehat{q}_{\mathcal{R}}^{(*)}(\alpha)) \le
        \min_{1\le t\le T} \ell(q^{(t)},\widehat{q}_{\mathcal{R}}^{(t)}(\alpha)) + \Delta \\ 
        + 3 L_W L_{[T]} \left[\left(\frac{\log(4T/\delta)}{N_{\mathcal{C}}}\right)^{\gamma/2} + \left(\frac{2}{N_{\mathcal{C}}}\right)^{\gamma}\right],        
    \end{gathered}
    \end{equation}
    where $\gamma$, $L_W$, and $L_{[T]} = \max_{1\le t\le T} L_t$ are constants defined in Assumptions \ref{assump:opt_technical_1} and \ref{assump:opt_technical_2}.
\end{theorem}


Again, in any setting where it is possible to sample from $\mathcal{P}(X, \Theta)$, $N_{\mathcal{C}}$ can be made arbitrarily large, tightening the bound in Equation \ref{eqn:iterate_thm}. 
Practitioners can, thus, focus on obtaining efficient predictors knowing that \textsc{CANVI} will make the optimal selection with high probability.

\begin{figure*}
\centering
\includegraphics[scale=0.23]{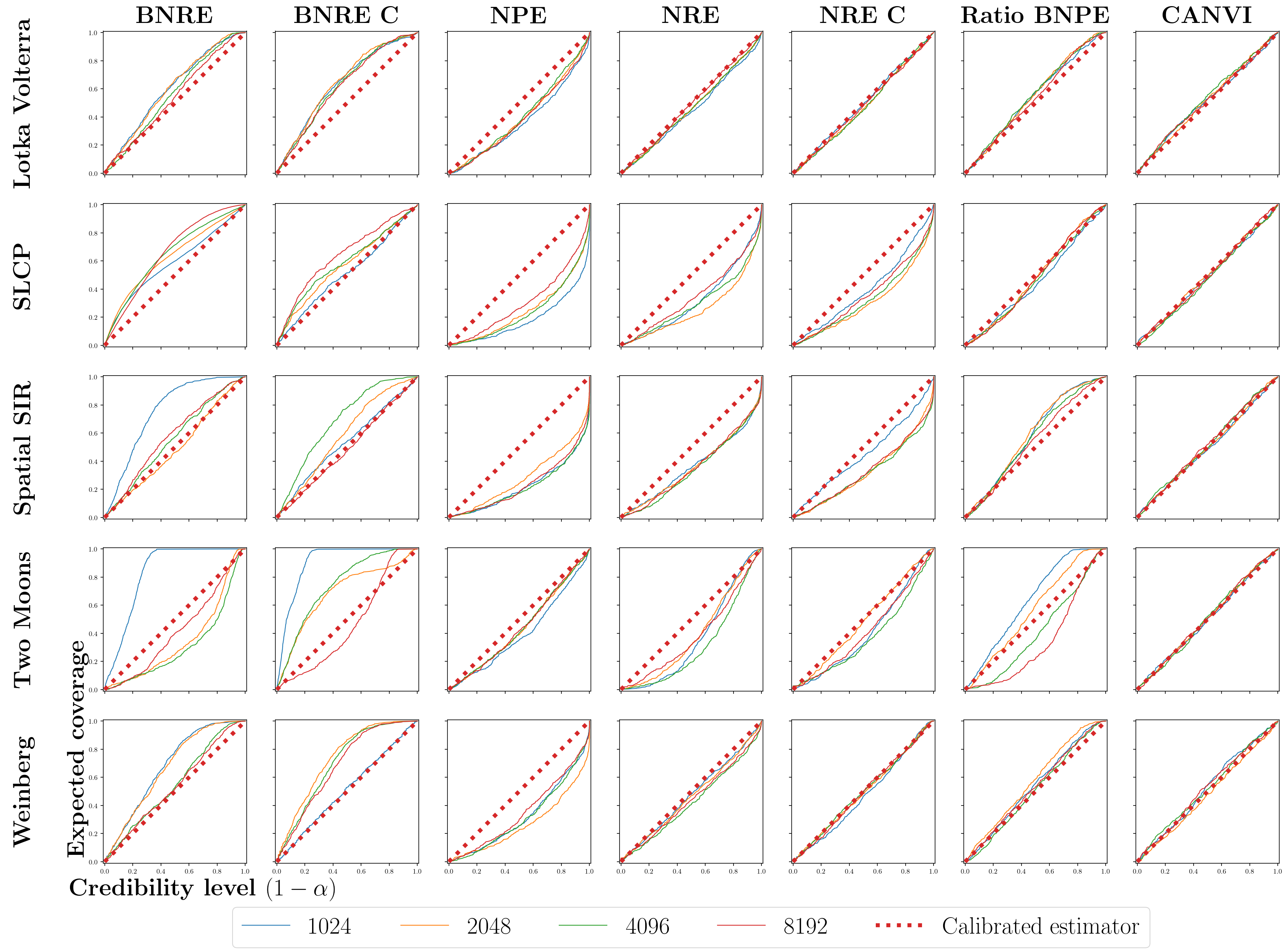}
\caption{\label{fig:coverage} Calibration on the SBI benchmarks across different calibration strategies. Perfect calibration corresponds to the highlighted $y=x$ curve. Conservative prediction regions lie above this calibrated line and overconfident ones below. Conformalized lines (\textsc{CANVI}) are difficult to distinguish, as they all lie along the desired $y=x$ curve. 
}
\end{figure*}

\section{Experiments}\label{section:experiments}
As discussed, the main advantages of \textsc{CANVI} over alternative strategies are its guaranteed calibration without the need for tuning parameters, minimal computational overhead, and informative prediction regions. For this reason, we present three experiments herein. The first (\Cref{section:coverage_comparison}) seeks to validate the first two claims by comparing BNRE, BNRE C, NPE, NRE, NRE C, and Ratio BNPE to the vanilla version of \textsc{CANVI} i.e. when it is applied to a single $q(\Theta\mid X)$, where no recalibration is necessary. 

Notably, the informativeness of prediction regions, the focus of the third claim, is only of interest once a predictor is calibrated, meaning the comparison with alternatives is only meaningful when they are nearly calibrated. As demonstrated in \Cref{section:coverage_comparison}, however, the alternatives fail to consistently demonstrate calibration, rendering such a comparison moot. For this reason, the following experiment (\Cref{section:empirical_eff}) focuses on demonstrating that applying the full \textsc{CANVI} procedure to a collection $\{q^{(t)}(\Theta\mid X)\}_{t=1}^T$ both retains coverage under recalibration and ultimately recovers the most efficient predictor. The latter hinges upon the validity of Assumption \ref{assump:opt_technical_3} for the Monte Carlo efficiency estimator in \Cref{eqn:cp_mc_est}. We, therefore, demonstrate in the subsequent experiments that this estimator exhibits the desired estimation consistency when applied in settings where posterior estimators are trained to different epochs (\Cref{section:eff_over_epochs}) or with different training objectives (\Cref{section:training_obj}). Finally, we demonstrate \textsc{CANVI} can computationally scale up to scientific problems of interest in \Cref{subsection:seds}.

In all experiments, coverage for variational posteriors is assessed using Monte Carlo estimation, namely by constructing the highest density credible region per $x_i$. That is, for a given $x_i$, the $\zeta$ such that $\{\theta_j \mid q(\theta_j\mid x_i) \ge \zeta\}$ captures $1-\alpha$ of the probability mass is estimated by drawing $\{\theta_j\}_{j=1}^N \sim q(\Theta | x_i)$ and finding the $1-\alpha$ quantile of $\{q(\theta_j| x_i)\}_{j=1}^N$. Coverage of the true parameter $\theta$ can be assessed by checking if $q(\theta\mid x_i)\ge\zeta$. Details are provided in Appendix \ref{section:exp_details}, and code is available at  \url{https://github.com/yashpatel5400/canvi.git}. 


\subsection{Coverage Calibration}\label{section:coverage_comparison}
We evaluate on the standard SBI benchmark tasks, highlighted in \citep{delaunoy2023balancing}.
For full descriptions of the tasks, refer to Appendix \ref{section:benchmark}. Again following the precedent from previous SBI works, we present the calibration of the models trained over several simulation budgets ($|\mathcal{D}|$). 
\textsc{CANVI} was applied to an NPE, in which $\mathcal{D}_C$ was taken to be 10\% of the simulation budgets and the remainder used for training. Calibration was completed in \textbf{under one second} for each task. Coverage was assessed 1,000 i.i.d. samples. Figure \ref{fig:coverage} demonstrates the miscalibration of the alternative approaches and the correction afforded with \textsc{CANVI}.

\subsection{Predictive Efficiency}\label{section:empirical_eff}
\subsubsection{Training Epochs}\label{section:eff_over_epochs}
We now study the application of \textsc{CANVI} to a collection of posteriors. In this experiment, $q^{(t)}$ is taken to be the $t$-th iterate of training a Neural Spline Flow family against the $\mathcal{L}_{\text{FAVI}}$ objective \citep{durkan2019neural}. Generally, we expect efficiency to improve with training iterates, as $q(\Theta\mid x)$ better approximates $\mathcal{P}(\Theta\mid x)$; however, the most efficient iterate may not occur at $t=T$. Selecting an intermediate $t$ is comparable to the practice of retaining the training iterate with the best validation performance in prediction tasks.

\begin{figure}[H]
\centering
\includegraphics[scale=0.18]{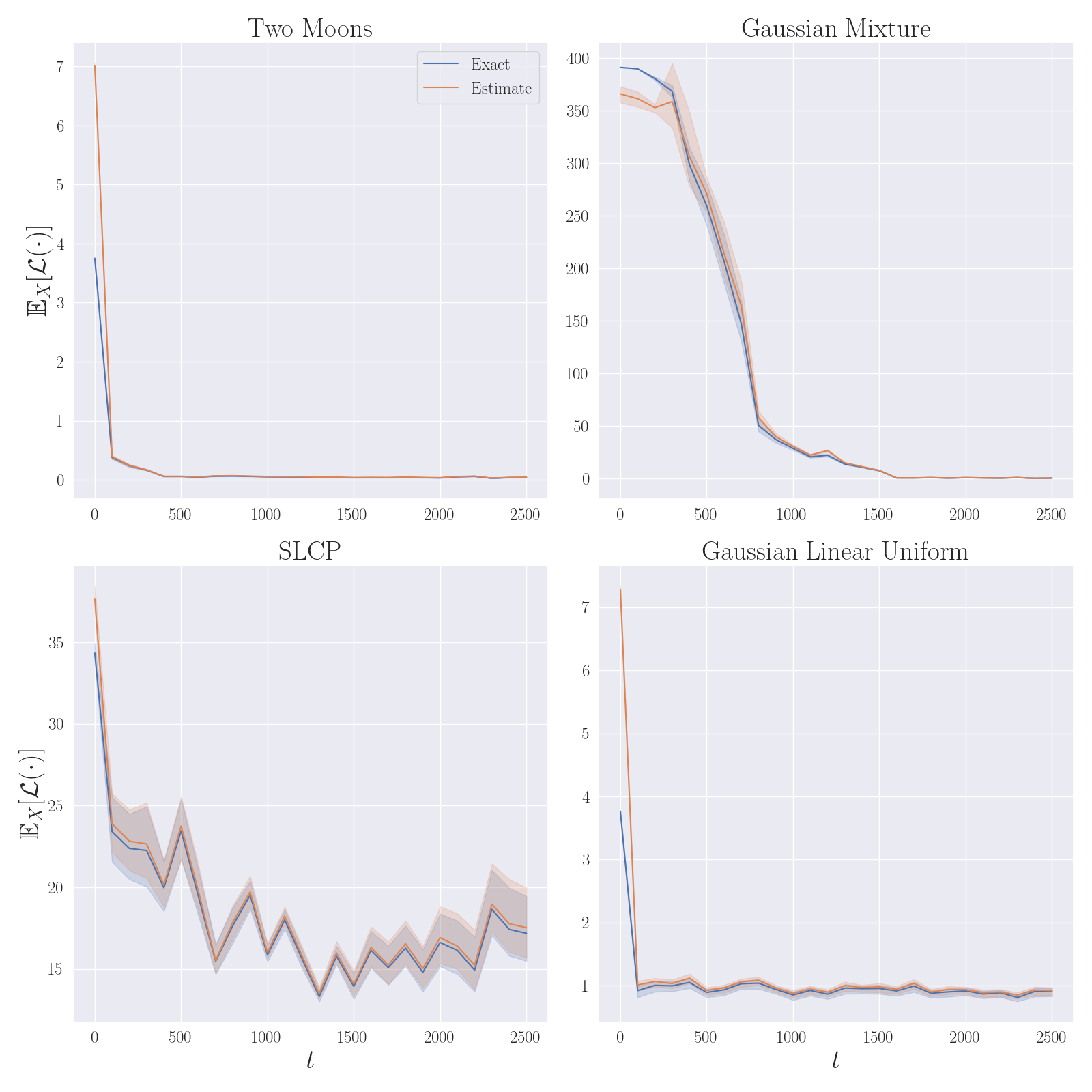}
\caption{\label{fig:efficiency_vs_training} $\ell(q,\tau)$ and $\widehat{\ell}(q, \tau)$ for $\tau = \widehat{q}_\mathcal{C}^{(t)}(\alpha)$.
Error bars across test batches are plotted.}
\end{figure}

\begin{table*}[t]
\caption{\label{table:arch_coverage} Coverage rates and standard errors for $\theta$ before (rows 1-4) and after conformalization (rows 5-8) by \textsc{CANVI}, for ARCH (left table) and SED (right table), assessed by checking for inclusion of $\theta$ in the $1-\alpha$ highest density region. Non-conformalized regions were estimated empirically from batches of 1000 i.i.d. samples per point. $\ell(q, \widehat{q}_{\mathcal{C}}(0.05))$ were computed with explicit gridding, discretizing each dimension into $200$ bins; such estimation was intractable for $\Theta\in\mathbb{R}^{11}$ for SEDs. $\widehat{\ell}$ was estimated using Algorithm \ref{alg:cp_region_size}, with $K=1$ and $K=10$ mixing discretizations.}
\resizebox{\textwidth}{!}{%

\begin{tabular}{cccc}
\toprule
$1-\alpha$ & ELBO & IWBO & FAVI \\
\midrule
0.50 & 0.0007 (0.0008) & 0.1031 (0.0110) & 0.5514 (0.0127) \\
0.75 & 0.0044 (0.0018) & 0.1994 (0.0115) & 0.7534 (0.0127) \\
0.90 & 0.0195 (0.0044) & 0.3263 (0.0122) & 0.8797 (0.0065) \\
0.95 & 0.0396 (0.0061) & 0.4074 (0.0186) & 0.9260 (0.0083) \\
\midrule
0.50 & 0.4970 (0.0124) & 0.5019 (0.0167) & 0.5086 (0.0144) \\
0.75 & 0.7488 (0.0139) & 0.7559 (0.0126) & 0.7565 (0.0092) \\
0.90 & 0.8978 (0.0080) & 0.9036 (0.0111) & 0.9005 (0.0110) \\
0.95 & 0.9496 (0.0071) & 0.9548 (0.0052) & 0.9487 (0.0081) \\
\bottomrule
$\ell(q, \widehat{q}_{\mathcal{C}}(0.05))$ & 1.3184 (0.2178) & 1.4211 (0.1128) & 0.7451 (0.0641) \\
$\widehat{\ell}_{K=1}(q, \widehat{q}_{\mathcal{C}}(0.05))$ & 50.6595 (8.1313) & 69.6484 (2.2495) & 19.3635 (0.8868) \\
$\widehat{\ell}_{K=10}(q, \widehat{q}_{\mathcal{C}}(0.05))$ & 1.4151 (0.2181) & 1.5206 (0.1114) & 0.8402 (0.0656) \\
\bottomrule
\end{tabular}
    \quad
\begin{tabular}{cccc}
\toprule
$1-\alpha$ & ELBO & IWBO & FAVI \\
\midrule
0.50 & 0.1683 (0.0088) & 0.6185 (0.0111) & 0.4986 (0.0180) \\
0.75 & 0.4556 (0.0146) & 0.7978 (0.0079) & 0.7550 (0.0105) \\
0.90 & 0.5803 (0.0106) & 0.8881 (0.0064) & 0.9028 (0.0092) \\
0.95 & 0.6824 (0.0144) & 0.9248 (0.0083) & 0.9532 (0.0085) \\
\midrule
0.50 & 0.4901 (0.0156) & 0.4937 (0.0101) & 0.5024 (0.0203) \\
0.75 & 0.7485 (0.0163) & 0.7513 (0.0139) & 0.7542 (0.0163) \\
0.90 & 0.8985 (0.0098) & 0.9013 (0.0096) & 0.9022 (0.0067) \\
0.95 & 0.9499 (0.0084) & 0.9510 (0.0061) & 0.9461 (0.0059) \\
\bottomrule
$\widehat{\ell}(q, \widehat{q}_{\mathcal{C}}(0.05))$ & $\infty$ & 1.3849 (1.2357) $\times10^{9}$ & 5.3732 (3.3302) $\times10^{6}$ \\
\bottomrule
\end{tabular}
}
\end{table*}

We first study the validity of Assumption \ref{assump:opt_technical_3} by comparing $\widehat{\ell}(q,\tau)$ to $\ell(q,\tau)$.
To compare to $\ell(q,\tau)$,
we restrict experiments to
projected posteriors of $\tilde{\Theta} = (\theta_1,\theta_2)$ for the Two Moons, Gaussian Mixture, SLCP, and Gaussian Linear Uniform tasks, for which explicit gridding was tractable.
$\widehat{\ell}(q^{(t)}, \widehat{q}_\mathcal{C}^{(t)}(\alpha))$ was estimated for a fixed $\alpha = 0.05$ and $t$ taken every 100 training steps with 5 batches of 100 test points ($|\mathcal{D_T}| = 100$) using $S=10,000$ importance-weighted i.i.d. samples for each of $K=10$ mixed samplers. 
From \Cref{fig:efficiency_vs_training}, we see that $\widehat{\ell}(q,\tau)$ tracks closely to $\ell(q,\tau)$ across all tasks, giving credence to Assumption \ref{assump:opt_technical_3}.
We visualize the credible regions in Appendix \ref{section:credible_regions}. 

\subsubsection{Training Objectives}\label{section:training_obj}



We now similarly study $q^{(t)}$ across training objectives, taking one iterate of $q$ trained against each of $\mathcal{L}_{\text{FAVI}}$, $\mathcal{L}_{\text{ELBO}}$, and $\mathcal{L}_{\text{IWBO}}$ ($K=10$), giving us three amortized posteriors as input for \textsc{CANVI}, for a lag-one ARCH model: 
\begin{equation*}
\label{arch1}
    y^{(m)}=\theta_1 y^{(m-1)}+e^{(m)}, \text{ } e^{(m)}=\xi^{(m)} \sqrt{0.2+\theta_2\left(e^{(m-1)}\right)^2},
\end{equation*}
where $y^{(0)}=0$, $e^{(0)}=0$, $M=100$, and the $\xi^{(m)}$ are independent standard normal random variables \citep{owen2022lfire}, detailed in Appendix \ref{section:arch_details}.

\Cref{table:arch_coverage} shows that prior to conformalization, the variational posteriors are generally miscalibrated: training by the ELBO or IWBO results in significant under-coverage, as targeting either is known to find solutions that are mode-seeking. While the variational posterior obtained by FAVI is better calibrated, it still needs correction. As multiple posterior approximators were considered, \textsc{CANVI} had to be applied with recalibration. 
\Cref{table:arch_coverage} shows that the recalibrated $1-\alpha$ prediction regions are nearly perfectly calibrated. Importantly, correction by \textsc{CANVI} can result in either larger or smaller $1-\alpha$ regions, depending on the direction of miscalibration. In settings where the variational posterior is overdispersed, applying \textsc{CANVI} results in smaller $1-\alpha$ density regions, explicitly shown in Appendix \ref{section:arch_post}. \Cref{table:arch_coverage} also demonstrates using the better calibrated $\mathcal{L}_{\text{FAVI}}$-trained approximation results in higher efficiency compared to the $\mathcal{L}_{\text{ELBO}}$ and $\mathcal{L}_{\text{IWBO}}$ counterparts. 

We also demonstrate the necessity of using mixed sampling for $\widehat{\ell}$. In particular, we compare the estimates when using only the variational posterior as a sampler ($\widehat{\ell}_{K=1}$) and when averaging 10 mixed samplers ($\widehat{\ell}_{K=10}$), from which we observe the estimator accuracy greatly improves with mixing, especially in the underdispersed IWBO and ELBO cases.

\subsection{Galaxy Spectral Energy Distributions}
\label{subsection:seds}

We now present the application of \textsc{CANVI} to an important scientific problem. The \textrm{spectrum} of an astronomical object is measured via a spectrograph, which records the flux across a large grid of wavelength values \citep{york2000sloan, abareshi2022overview}, which we simulate with the Probabilistic Value-Added Bright Galaxy Survey simulator (PROVABGS). PROVABGS
maps $\theta\in\mathbb{R}^{11}$ to galaxy spectra, detailed further and visualized in Appendix \ref{section:sed_setup}.

A mixture of 20 Gaussian distributions was used as the variational posterior and trained against the $\mathcal{L}_{\text{FAVI}},\mathcal{L}_{\text{ELBO}},$ and $\mathcal{L}_{\text{IWBO}}$ objectives, as in \Cref{section:training_obj}. \Cref{table:arch_coverage} shows that the ELBO and IWBO tend to be overly concentrated, failing to contain the entire parameter vector $\theta$ in the $1-\alpha$ highest-density region often. FAVI, on the other hand, is reasonably well-calibrated. After applying \textsc{CANVI}, all three methods achieve nearly perfect calibration across a range of desired confidence levels. Of course, the utility of these corrected regions depends on the level of information contained in the original model. For the ELBO or IWBO cases, the corrected regions achieve statistical validity, but are too large to be informative. Notably, the underdispersion of the ELBO approximator led to $\widehat{q}_{\mathcal{C}}(0.05) = 0$ (up to machine precision), resulting in a volume estimate of $\infty$. For FAVI, on the other hand, application of \textsc{CANVI} results in statistical guarantees with minimal alterations to the high-density regions.

\section{Discussion}\label{section:discussion} 
We have presented \textsc{CANVI}, a framework for producing marginally calibrated, efficient prediction regions from a collection of variational approximators with minimal overhead in simulation-based inference settings.
We view such guarantees on marginal coverage as an important first step toward increasing the utility of such inference algorithms for downstream applications, suggesting many directions for extension. Of immediate interest would be extending \textsc{CANVI} to cases of misspecified forward models $\mathcal{P}(X\mid \Theta)$ by leveraging work on conformal prediction under distribution shift \citep{tibshirani2019conformal,barber2022conformal}. 
Further, leveraging \textsc{CANVI} over functional spaces may enable guarantees over the full posterior distributions. 


\section*{Impact Statement}
This paper presents work whose goal is to advance the field of Machine Learning. There are many potential societal consequences of our work, none which we feel must be specifically highlighted here.


\clearpage
\bibliography{example_paper}

\begin{thebibliography}{51}
\providecommand{\natexlab}[1]{#1}
\providecommand{\url}[1]{\texttt{#1}}
\expandafter\ifx\csname urlstyle\endcsname\relax
  \providecommand{\doi}[1]{doi: #1}\else
  \providecommand{\doi}{doi: \begingroup \urlstyle{rm}\Url}\fi

\bibitem[Abareshi et~al.(2022)Abareshi, Aguilar, Ahlen, Alam, Alexander, Alfarsy, Allen, Prieto, Alves, Ameel, et~al.]{abareshi2022overview}
Abareshi, B., Aguilar, J., Ahlen, S., Alam, S., Alexander, D.~M., Alfarsy, R., Allen, L., Prieto, C.~A., Alves, O., Ameel, J., et~al.
\newblock Overview of the instrumentation for the {Dark Energy Spectroscopic Instrument}.
\newblock \emph{The Astronomical Journal}, 164\penalty0 (5):\penalty0 207, 2022.

\bibitem[Agrawal et~al.(2020)Agrawal, Sheldon, and Domke]{agrawal2020advances}
Agrawal, A., Sheldon, D.~R., and Domke, J.
\newblock Advances in black-box {VI}: Normalizing flows, importance weighting, and optimization.
\newblock \emph{Advances in Neural Information Processing Systems}, 33:\penalty0 17358--17369, 2020.

\bibitem[Ambrogioni et~al.(2019)Ambrogioni, G{\"u}{\c{c}}l{\"u}, Berezutskaya, Borne, G{\"u}{\c{c}}l{\"u}t{\"u}rk, Hinne, Maris, and Gerven]{ambrogioni2019forward}
Ambrogioni, L., G{\"u}{\c{c}}l{\"u}, U., Berezutskaya, J., Borne, E., G{\"u}{\c{c}}l{\"u}t{\"u}rk, Y., Hinne, M., Maris, E., and Gerven, M.
\newblock Forward amortized inference for likelihood-free variational marginalization.
\newblock In \emph{The 22nd International Conference on Artificial Intelligence and Statistics}, pp.\  777--786. PMLR, 2019.

\bibitem[Angelopoulos \& Bates(2021)Angelopoulos and Bates]{angelopoulos2021gentle}
Angelopoulos, A.~N. and Bates, S.
\newblock A gentle introduction to conformal prediction and distribution-free uncertainty quantification.
\newblock \emph{arXiv preprint arXiv:2107.07511}, 2021.

\bibitem[Barber et~al.(2022)Barber, Candes, Ramdas, and Tibshirani]{barber2022conformal}
Barber, R.~F., Candes, E.~J., Ramdas, A., and Tibshirani, R.~J.
\newblock Conformal prediction beyond exchangeability.
\newblock \emph{arXiv preprint arXiv:2202.13415}, 2022.

\bibitem[Blei et~al.(2017)Blei, Kucukelbir, and McAuliffe]{blei2017variational}
Blei, D.~M., Kucukelbir, A., and McAuliffe, J.~D.
\newblock Variational inference: A review for statisticians.
\newblock \emph{Journal of the American statistical Association}, 112\penalty0 (518):\penalty0 859--877, 2017.

\bibitem[Boelts et~al.(2022)Boelts, Lueckmann, Gao, and Macke]{boelts2022flexible}
Boelts, J., Lueckmann, J.-M., Gao, R., and Macke, J.~H.
\newblock Flexible and efficient simulation-based inference for models of decision-making.
\newblock \emph{Elife}, 11:\penalty0 e77220, 2022.

\bibitem[Bornschein \& Bengio(2014)Bornschein and Bengio]{bornschein2014reweighted}
Bornschein, J. and Bengio, Y.
\newblock Reweighted wake-sleep.
\newblock \emph{arXiv preprint arXiv:1406.2751}, 2014.

\bibitem[Burda et~al.(2015)Burda, Grosse, and Salakhutdinov]{burda2015importance}
Burda, Y., Grosse, R., and Salakhutdinov, R.
\newblock Importance weighted autoencoders.
\newblock \emph{arXiv preprint arXiv:1509.00519}, 2015.

\bibitem[Cranmer et~al.(2020)Cranmer, Brehmer, and Louppe]{cranmer2020frontier}
Cranmer, K., Brehmer, J., and Louppe, G.
\newblock The frontier of simulation-based inference.
\newblock \emph{Proceedings of the National Academy of Sciences}, 117\penalty0 (48):\penalty0 30055--30062, 2020.

\bibitem[Deistler et~al.(2022)Deistler, Goncalves, and Macke]{deistler2022truncated}
Deistler, M., Goncalves, P.~J., and Macke, J.~H.
\newblock Truncated proposals for scalable and hassle-free simulation-based inference.
\newblock \emph{arXiv preprint arXiv:2210.04815}, 2022.

\bibitem[Delaunoy et~al.(2022)Delaunoy, Hermans, Rozet, Wehenkel, and Louppe]{delaunoy2022towards}
Delaunoy, A., Hermans, J., Rozet, F., Wehenkel, A., and Louppe, G.
\newblock Towards reliable simulation-based inference with balanced neural ratio estimation.
\newblock \emph{arXiv preprint arXiv:2208.13624}, 2022.

\bibitem[Delaunoy et~al.(2023)Delaunoy, Miller, Forr{\'e}, Weniger, and Louppe]{delaunoy2023balancing}
Delaunoy, A., Miller, B.~K., Forr{\'e}, P., Weniger, C., and Louppe, G.
\newblock Balancing simulation-based inference for conservative posteriors.
\newblock \emph{arXiv preprint arXiv:2304.10978}, 2023.

\bibitem[Durkan et~al.(2019)Durkan, Bekasov, Murray, and Papamakarios]{durkan2019neural}
Durkan, C., Bekasov, A., Murray, I., and Papamakarios, G.
\newblock Neural spline flows.
\newblock \emph{Advances in neural information processing systems}, 32, 2019.

\bibitem[Durkan et~al.(2020{\natexlab{a}})Durkan, Bekasov, Murray, and Papamakarios]{nflows}
Durkan, C., Bekasov, A., Murray, I., and Papamakarios, G.
\newblock {nflows}: normalizing flows in {PyTorch}, November 2020{\natexlab{a}}.
\newblock URL \url{https://doi.org/10.5281/zenodo.4296287}.

\bibitem[Durkan et~al.(2020{\natexlab{b}})Durkan, Murray, and Papamakarios]{durkan2020contrastive}
Durkan, C., Murray, I., and Papamakarios, G.
\newblock On contrastive learning for likelihood-free inference.
\newblock In \emph{International conference on machine learning}, pp.\  2771--2781. PMLR, 2020{\natexlab{b}}.

\bibitem[Falkiewicz et~al.(2023)Falkiewicz, Takeishi, Shekhzadeh, Wehenkel, Delaunoy, Louppe, and Kalousis]{falkiewicz2023calibrating}
Falkiewicz, M., Takeishi, N., Shekhzadeh, I., Wehenkel, A., Delaunoy, A., Louppe, G., and Kalousis, A.
\newblock Calibrating neural simulation-based inference with differentiable coverage probability.
\newblock \emph{arXiv preprint arXiv:2310.13402}, 2023.

\bibitem[Gilman et~al.(2021)Gilman, Bovy, Treu, Nierenberg, Birrer, Benson, and Sameie]{gilman2021strong}
Gilman, D., Bovy, J., Treu, T., Nierenberg, A., Birrer, S., Benson, A., and Sameie, O.
\newblock Strong lensing signatures of self-interacting dark matter in low-mass haloes.
\newblock \emph{Monthly Notices of the Royal Astronomical Society}, 507\penalty0 (2):\penalty0 2432--2447, 2021.

\bibitem[Greenberg et~al.(2019)Greenberg, Nonnenmacher, and Macke]{greenberg2019automatic}
Greenberg, D., Nonnenmacher, M., and Macke, J.
\newblock Automatic posterior transformation for likelihood-free inference.
\newblock In \emph{International Conference on Machine Learning}, pp.\  2404--2414. PMLR, 2019.

\bibitem[Gupta et~al.(2022)Gupta, Kuchibhotla, and Ramdas]{gupta2022nested}
Gupta, C., Kuchibhotla, A.~K., and Ramdas, A.
\newblock Nested conformal prediction and quantile out-of-bag ensemble methods.
\newblock \emph{Pattern Recognition}, 127:\penalty0 108496, 2022.

\bibitem[Hahn et~al.(2023)Hahn, Kwon, Tojeiro, Siudek, Canning, Mezcua, Tinker, Brooks, Doel, Fanning, et~al.]{hahn2023desi}
Hahn, C., Kwon, K., Tojeiro, R., Siudek, M., Canning, R.~E., Mezcua, M., Tinker, J.~L., Brooks, D., Doel, P., Fanning, K., et~al.
\newblock The {DESI PRObabilistic Value-added Bright Galaxy Survey (PROVABGS)} mock challenge.
\newblock \emph{The Astrophysical Journal}, 945\penalty0 (1):\penalty0 16, 2023.

\bibitem[Hermans et~al.(2020)Hermans, Begy, and Louppe]{hermans2020likelihood}
Hermans, J., Begy, V., and Louppe, G.
\newblock Likelihood-free mcmc with amortized approximate ratio estimators.
\newblock In \emph{International conference on machine learning}, pp.\  4239--4248. PMLR, 2020.

\bibitem[Hermans et~al.(2021{\natexlab{a}})Hermans, Banik, Weniger, Bertone, and Louppe]{hermans2021towards}
Hermans, J., Banik, N., Weniger, C., Bertone, G., and Louppe, G.
\newblock Towards constraining warm dark matter with stellar streams through neural simulation-based inference.
\newblock \emph{Monthly Notices of the Royal Astronomical Society}, 507\penalty0 (2):\penalty0 1999--2011, 2021{\natexlab{a}}.

\bibitem[Hermans et~al.(2021{\natexlab{b}})Hermans, Delaunoy, Rozet, Wehenkel, and Louppe]{hermans2021averting}
Hermans, J., Delaunoy, A., Rozet, F., Wehenkel, A., and Louppe, G.
\newblock Averting a crisis in simulation-based inference.
\newblock \emph{arXiv preprint arXiv:2110.06581}, 2021{\natexlab{b}}.

\bibitem[Hezaveh et~al.(2016)Hezaveh, Dalal, Marrone, Mao, Morningstar, Wen, Blandford, Carlstrom, Fassnacht, Holder, et~al.]{hezaveh2016detection}
Hezaveh, Y.~D., Dalal, N., Marrone, D.~P., Mao, Y.-Y., Morningstar, W., Wen, D., Blandford, R.~D., Carlstrom, J.~E., Fassnacht, C.~D., Holder, G.~P., et~al.
\newblock Detection of lensing substructure using {ALMA} observations of the dusty galaxy {SDP.81}.
\newblock \emph{The Astrophysical Journal}, 823\penalty0 (1):\penalty0 37, 2016.

\bibitem[Hogg \& Blandford(1994)Hogg and Blandford]{hogg1994gravitational}
Hogg, D.~W. and Blandford, R.
\newblock The gravitational lens system {B1422+231}: dark matter, superluminal expansion and the {Hubble} constant.
\newblock \emph{Monthly Notices of the Royal Astronomical Society}, 268\penalty0 (4):\penalty0 889--893, 1994.

\bibitem[Jubin(2019)]{jubin2019intrinsic}
Jubin, B.
\newblock Intrinsic volumes of sublevel sets.
\newblock \emph{arXiv preprint arXiv:1903.01592}, 2019.

\bibitem[Kingma \& Ba(2014)Kingma and Ba]{kingma2014adam}
Kingma, D.~P. and Ba, J.
\newblock Adam: A method for stochastic optimization.
\newblock \emph{arXiv preprint arXiv:1412.6980}, 2014.

\bibitem[K{\"o}hler et~al.(2021)K{\"o}hler, Kr{\"a}mer, and No{\'e}]{kohler2021smooth}
K{\"o}hler, J., Kr{\"a}mer, A., and No{\'e}, F.
\newblock Smooth normalizing flows.
\newblock \emph{Advances in Neural Information Processing Systems}, 34:\penalty0 2796--2809, 2021.

\bibitem[Lee \& Lee(2012)Lee and Lee]{lee2012smooth}
Lee, J.~M. and Lee, J.~M.
\newblock \emph{Smooth manifolds}.
\newblock Springer, 2012.

\bibitem[Lei et~al.(2018)Lei, G’Sell, Rinaldo, Tibshirani, and Wasserman]{lei2018distribution}
Lei, J., G’Sell, M., Rinaldo, A., Tibshirani, R.~J., and Wasserman, L.
\newblock Distribution-free predictive inference for regression.
\newblock \emph{Journal of the American Statistical Association}, 113\penalty0 (523):\penalty0 1094--1111, 2018.

\bibitem[Lemos et~al.(2023)Lemos, Coogan, Hezaveh, and Perreault-Levasseur]{lemos2023sampling}
Lemos, P., Coogan, A., Hezaveh, Y., and Perreault-Levasseur, L.
\newblock Sampling-based accuracy testing of posterior estimators for general inference.
\newblock \emph{arXiv preprint arXiv:2302.03026}, 2023.

\bibitem[Lueckmann et~al.(2017)Lueckmann, Goncalves, Bassetto, {\"O}cal, Nonnenmacher, and Macke]{lueckmann2017flexible}
Lueckmann, J.-M., Goncalves, P.~J., Bassetto, G., {\"O}cal, K., Nonnenmacher, M., and Macke, J.~H.
\newblock Flexible statistical inference for mechanistic models of neural dynamics.
\newblock \emph{Advances in neural information processing systems}, 30, 2017.

\bibitem[Lueckmann et~al.(2021)Lueckmann, Boelts, Greenberg, Goncalves, and Macke]{lueckmann2021benchmarking}
Lueckmann, J.-M., Boelts, J., Greenberg, D., Goncalves, P., and Macke, J.
\newblock Benchmarking simulation-based inference.
\newblock In \emph{International Conference on Artificial Intelligence and Statistics}, pp.\  343--351. PMLR, 2021.

\bibitem[Miller et~al.(2022)Miller, Weniger, and Forr{\'e}]{miller2022contrastive}
Miller, B.~K., Weniger, C., and Forr{\'e}, P.
\newblock Contrastive neural ratio estimation.
\newblock \emph{arXiv preprint arXiv:2210.06170}, 2022.

\bibitem[Murphy(2022)]{murphy2022probabilistic}
Murphy, K.~P.
\newblock \emph{Probabilistic machine learning: an introduction}.
\newblock MIT press, 2022.

\bibitem[Murphy(2023)]{pml2Book}
Murphy, K.~P.
\newblock \emph{Probabilistic Machine Learning: Advanced Topics}.
\newblock MIT Press, 2023.
\newblock URL \url{probml.ai}.

\bibitem[Papamakarios \& Murray(2016)Papamakarios and Murray]{papamakarios2016fast}
Papamakarios, G. and Murray, I.
\newblock Fast $\varepsilon$-free inference of simulation models with bayesian conditional density estimation.
\newblock \emph{Advances in neural information processing systems}, 29, 2016.

\bibitem[Papamakarios et~al.(2019)Papamakarios, Sterratt, and Murray]{papamakarios2019sequential}
Papamakarios, G., Sterratt, D., and Murray, I.
\newblock Sequential neural likelihood: Fast likelihood-free inference with autoregressive flows.
\newblock In \emph{The 22nd International Conference on Artificial Intelligence and Statistics}, pp.\  837--848. PMLR, 2019.

\bibitem[Paszke et~al.(2019)Paszke, Gross, Massa, Lerer, Bradbury, Chanan, Killeen, Lin, Gimelshein, Antiga, et~al.]{paszke2019pytorch}
Paszke, A., Gross, S., Massa, F., Lerer, A., Bradbury, J., Chanan, G., Killeen, T., Lin, Z., Gimelshein, N., Antiga, L., et~al.
\newblock Pytorch: An imperative style, high-performance deep learning library.
\newblock \emph{Advances in neural information processing systems}, 32, 2019.

\bibitem[Rezende \& Mohamed(2015)Rezende and Mohamed]{rezende2015variational}
Rezende, D. and Mohamed, S.
\newblock Variational inference with normalizing flows.
\newblock In \emph{International conference on machine learning}, pp.\  1530--1538. PMLR, 2015.

\bibitem[Sesia \& Cand{\`e}s(2020)Sesia and Cand{\`e}s]{sesia2020comparison}
Sesia, M. and Cand{\`e}s, E.~J.
\newblock A comparison of some conformal quantile regression methods.
\newblock \emph{Stat}, 9\penalty0 (1):\penalty0 e261, 2020.

\bibitem[Shafer \& Vovk(2008)Shafer and Vovk]{shafer2008tutorial}
Shafer, G. and Vovk, V.
\newblock A tutorial on conformal prediction.
\newblock \emph{Journal of Machine Learning Research}, 9\penalty0 (3), 2008.

\bibitem[Thomas et~al.(2022)Thomas, Dutta, Corander, Kaski, and Gutmann]{owen2022lfire}
Thomas, O., Dutta, R., Corander, J., Kaski, S., and Gutmann, M.~U.
\newblock {Likelihood-Free Inference by Ratio Estimation}.
\newblock \emph{Bayesian Analysis}, 17\penalty0 (1):\penalty0 1 -- 31, 2022.
\newblock \doi{10.1214/20-BA1238}.
\newblock URL \url{https://doi.org/10.1214/20-BA1238}.

\bibitem[Tibshirani et~al.(2019)Tibshirani, Foygel~Barber, Candes, and Ramdas]{tibshirani2019conformal}
Tibshirani, R.~J., Foygel~Barber, R., Candes, E., and Ramdas, A.
\newblock Conformal prediction under covariate shift.
\newblock \emph{Advances in neural information processing systems}, 32, 2019.

\bibitem[Vegetti \& Koopmans(2009)Vegetti and Koopmans]{vegetti2009statistics}
Vegetti, S. and Koopmans, L.
\newblock Statistics of mass substructure from strong gravitational lensing: quantifying the mass fraction and mass function.
\newblock \emph{Monthly Notices of the Royal Astronomical Society}, 400\penalty0 (3):\penalty0 1583--1592, 2009.

\bibitem[Vegetti et~al.(2010)Vegetti, Koopmans, Bolton, Treu, and Gavazzi]{vegetti2010detection}
Vegetti, S., Koopmans, L., Bolton, A., Treu, T., and Gavazzi, R.
\newblock Detection of a dark substructure through gravitational imaging.
\newblock \emph{Monthly Notices of the Royal Astronomical Society}, 408\penalty0 (4):\penalty0 1969--1981, 2010.

\bibitem[Yang \& Kuchibhotla(2021)Yang and Kuchibhotla]{yang2021finite}
Yang, Y. and Kuchibhotla, A.~K.
\newblock Finite-sample efficient conformal prediction.
\newblock \emph{arXiv preprint arXiv:2104.13871}, 2021.

\bibitem[Yao et~al.(2018)Yao, Vehtari, Simpson, and Gelman]{yao2018yes}
Yao, Y., Vehtari, A., Simpson, D., and Gelman, A.
\newblock Yes, but did it work?: Evaluating variational inference.
\newblock In \emph{International Conference on Machine Learning}, pp.\  5581--5590. PMLR, 2018.

\bibitem[York et~al.(2000)York, Adelman, Anderson~Jr, Anderson, Annis, Bahcall, Bakken, Barkhouser, Bastian, Berman, et~al.]{york2000sloan}
York, D.~G., Adelman, J., Anderson~Jr, J.~E., Anderson, S.~F., Annis, J., Bahcall, N.~A., Bakken, J., Barkhouser, R., Bastian, S., Berman, E., et~al.
\newblock The {Sloan Digital Sky Survey}: Technical summary.
\newblock \emph{The Astronomical Journal}, 120\penalty0 (3):\penalty0 1579, 2000.

\bibitem[Zhang et~al.(2018)Zhang, B{\"u}tepage, Kjellstr{\"o}m, and Mandt]{zhang2018advances}
Zhang, C., B{\"u}tepage, J., Kjellstr{\"o}m, H., and Mandt, S.
\newblock Advances in variational inference.
\newblock \emph{IEEE transactions on pattern analysis and machine intelligence}, 41\penalty0 (8):\penalty0 2008--2026, 2018.

\end{thebibliography}
\bibliographystyle{icml2024}

\newpage
\appendix
\onecolumn
\section{Group Conditional \textsc{CANVI}}\label{section:group_cond}
We now discuss how \textsc{CANVI} can be easily extended to provide a stronger notion of group conditional coverage over the purely marginal coverage over $\mathcal{X}$ that was provided in the paper. Here, instead of defining a global $\widehat{q}$, a collection $\{\widehat{q}_i\}_{i=1}^{K}$ is obtained by defining centroids over the $\mathcal{X}$ space $\{c_i\}_{i=1}^{K}$, constructing balls of radius $\epsilon$ around them $\mathcal{B}_{\epsilon}(c_i)\subset\mathcal{X}$, and performing calibration using $\mathcal{D}_{\mathcal{C}}^i := \{(x_j,\theta_j)\}_{j=1}^{\mathcal{N}_C}$ with $x_j\in\mathcal{B}_{\epsilon}(c_i)$. 

To compare, we define a \textit{global} quantile $\widehat{q}$ using $KN$ samples. Specifically, we take $N = 100,000$ and $K = 5$, meaning calibration was performed using 500,000 i.i.d. samples for the \textit{overall} calibration set. Coverage was assessed over $\alpha\in[0,1]$ discretized at steps of .05 over $K = 5$ regions. Coverage was assessed over 10 batches of 100,000 i.i.d. test samples. 

\begin{figure}[H]
\centering
\includegraphics[scale=0.24]{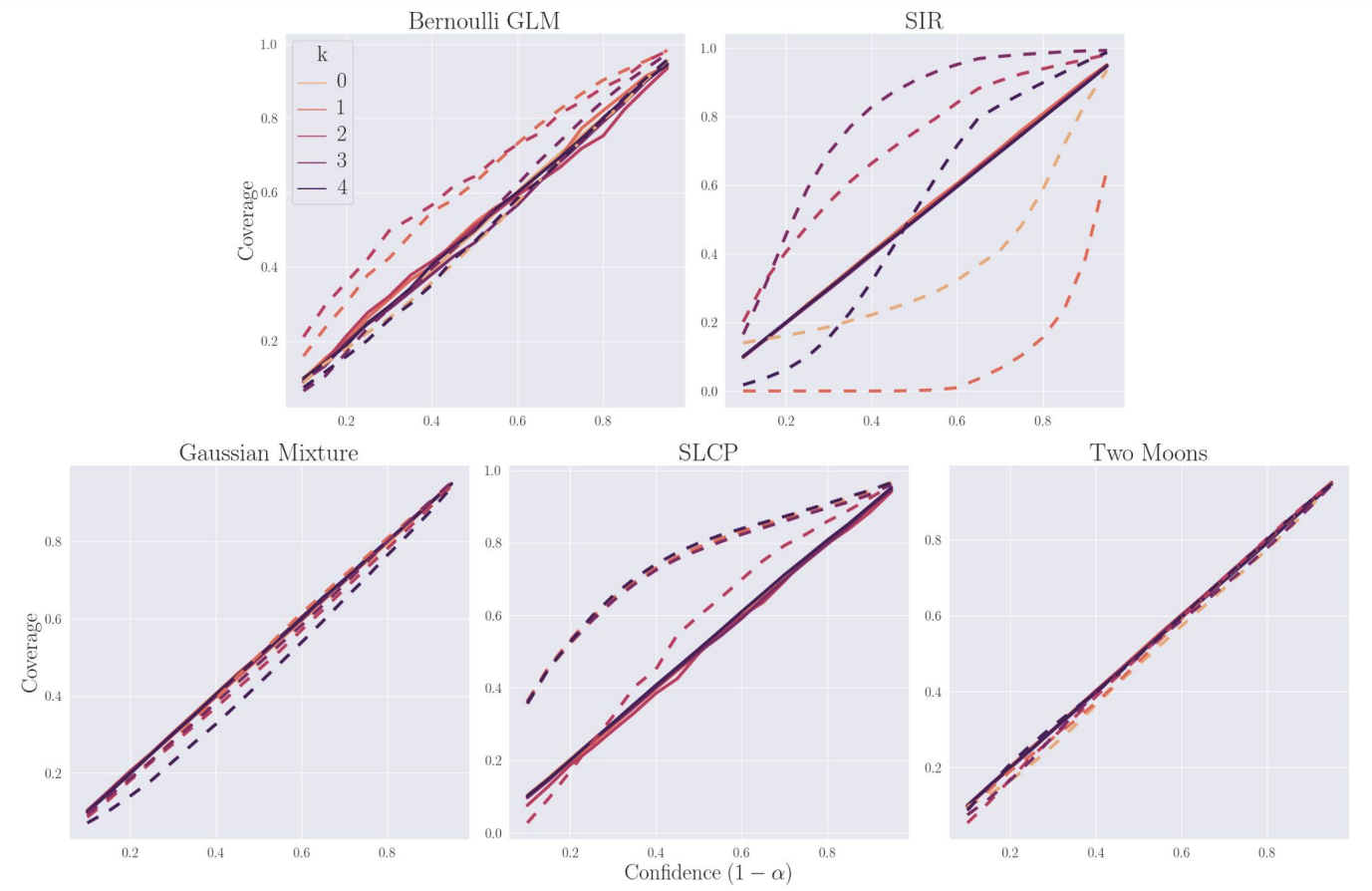}
\caption{\label{fig:group_cond_coverage} Calibration on the SBI benchmarks with overall quantile and region-specific quantiles, respectively the dashed and solid lines. Region-specific conformalized lines are slightly difficult to distinguish, as they all lie along the desired $y=x$ curve. Error bars from coverage assessments across test batches are plotted, although they are difficult to see due to the low variance between estimates across batches.}
\end{figure}

\begin{figure}[H]
\centering
\includegraphics[scale=0.27]{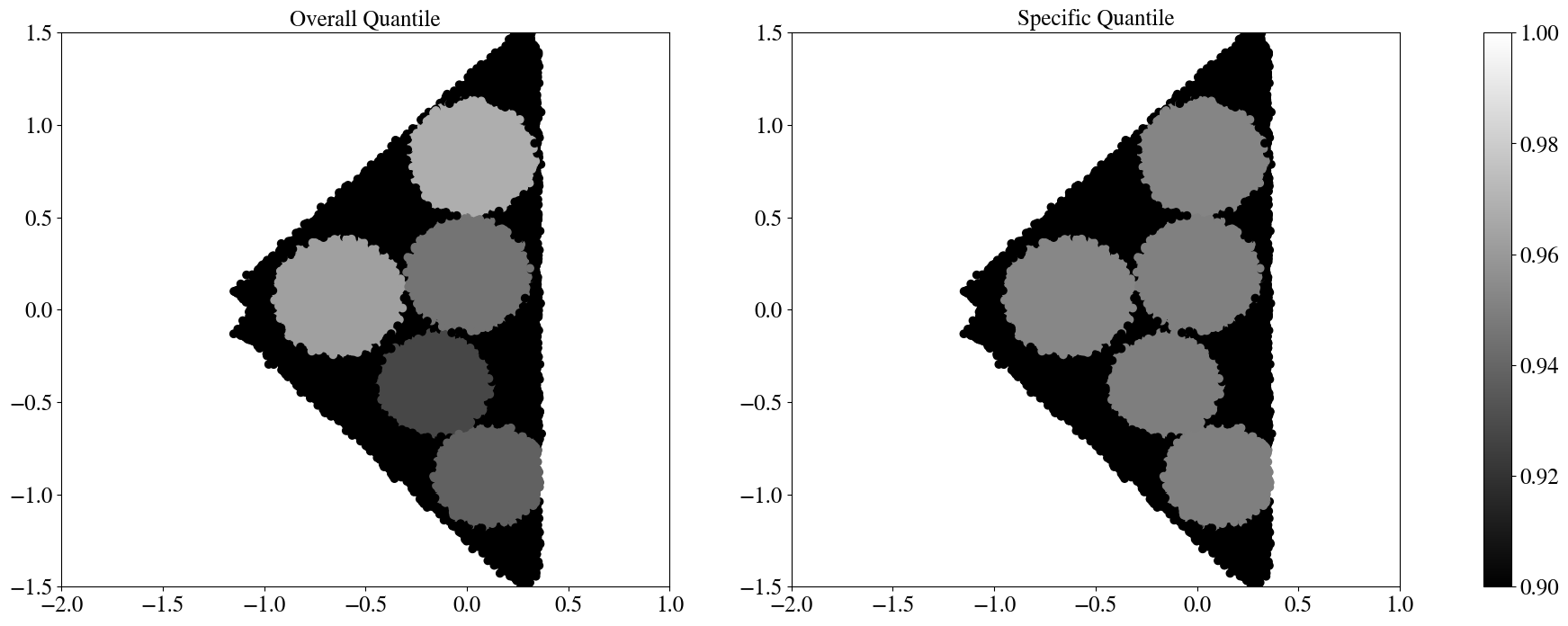}
\caption{\label{fig:overall_vs_cond} Coverage for Two Moons with the overall and region-specific quantiles with $\alpha=0.05$.}
\end{figure}

Figure \ref{fig:group_cond_coverage} demonstrates the miscalibration of the regions produced using the overall quantile and subsequent correction with region-specific calibration. We additionally plot the calibration across the $X$ space of the Two Moons task to demonstrate the non-uniformity of coverage across regions and subsequent corrections in Figure \ref{fig:overall_vs_cond}.

\section{CANVI Validity}\label{section:canvi_validity}
\begin{lemma}
    Let $\alpha\in(0,1)$ and
    \begin{gather*}
        q^{(*)}(\Theta\mid X),\widehat{q}_{\mathcal{R}}^{(*)}(\alpha) = \\
        \mathrm{CANVI}\left(
        \{q^{(t)}(\Theta\mid X)\}_{t=1}^T,\mathcal{P}(X, \Theta),1-\alpha,N_{\mathcal{C}},N_{\mathcal{T}}
    \right)
    \end{gather*}
    Let $(x', \theta')\sim\mathcal{P}(X, \Theta) \indep \mathcal{D} \cup \mathcal{D_C} \cup \mathcal{D_R} \cup \mathcal{D_T}$, with $\mathcal{D}$ being the data used to train $\{q^{(t)}(\Theta|X)\}_{t=1}^T$. Then $1 - \alpha\le\mathcal{P}(1/q^{(*)}(\theta'|x') \leq \widehat{q}_{\mathcal{R}}^{(*)}(\alpha))$.
\end{lemma}

\begin{proof}
Consider the score function $s^{(*)}(x, \theta) := 1 / q^{(*)}(\theta\mid x)$. 
Observe that $(x', \theta') \cup \mathcal{D_R}$ are jointly sampled i.i.d.~from $\mathcal{P}(X, \Theta)$, independent of the datasets used to design $s^{(*)}(x, \theta)$, namely $\mathcal{D} \cup \mathcal{D_C} \cup \mathcal{D_T}$. Denoting $\mathcal{S}_{\mathcal{R}} := \{s^{(*)}(x_i, \theta_i)\}_{(x_i,\theta_i)\in\mathcal{D_R}}$, scores $s^{(*)}(x', \theta') \cup \mathcal{S}_{\mathcal{R}}$, thus, too are i.i.d and, hence, exchangeable. The coverage guarantee then follows from the general theory of conformal prediction, presented in \citep{angelopoulos2021gentle}. 
\end{proof}

\section{Riemannian Manifolds}\label{section:riemannian_review}
Let $(\mathcal{M}, g)$ denote a Riemannian manifold, with $g$ denoting the metric tensor associated with this space. By definition, a manifold $\mathcal{M}$ is locally isomorphic to Euclidean space, from which we can define the notion of a tangent space $\mathcal{T}_z\mathcal{M}$ at each point $z\in\mathcal{M}$. The metric tensor $g$ then defines a \textit{local} notion of distance, namely $g : \mathcal{T}_z\mathcal{M}\times\mathcal{T}_z\mathcal{M}\rightarrow\mathbb{R}$. This, therefore, induces a local notion of length, namely for $x\in\mathcal{T}_z\mathcal{M},||x||_z = \sqrt{g(x,x)}$.

\textit{Global} distances over the manifold, therefore, can then be denoted by integrating such a local notion across a path $\gamma$. Concretely, a path is defined between $a,b\in\mathcal{M}$ by $\gamma : [0,1]\rightarrow\mathcal{M}$ such that $\gamma(0)=a,\gamma(1) = b$. The length of such a path is then $\ell(\gamma) := \int_{0}^1 || \gamma'(t) ||_{\gamma(t)} dt$. A natural choice of distance, therefore, is the minimum length of a constant velocity path, formally
\begin{equation}
    d(a,b) = \inf_{\gamma : || \gamma'(t) || = 1, \gamma(0)=a,\gamma(1)=b} \ell(\gamma)
\end{equation}

\section{CANVI: Efficiency Analysis}

\subsection{Lipschitz Continuity of Efficiency}\label{section:lip_cts_expectation}

\begin{lemma}
    Let $\ell(q, \tau)$ be as defined in Equation \ref{eqn:cp_global_eff}. If $\ell_{x}(q, \tau)$ is $L$-Lipschitz continuous in $\tau$ for any $x\in \mathcal{X}$, then $\ell(q, \tau)$ is $L$-Lipschitz continuous.
\end{lemma}

\begin{proof}
\begin{gather*}
    \left|\ell(q, \tau_1) - \ell(q, \tau_2)\right| 
    = \left|\mathbb{E}_X\left[\mathcal{L}\left(\left\{\theta : 1/ q(\Theta\mid X) \le \tau_1\right\}\right) - \mathcal{L}\left(\left\{\theta : 1/ q(\Theta\mid X) \le \tau_2\right\}\right)\right]\right| \\ 
    = \left|
    \int \mathcal{P}(x) \left[\ell_{x}(q, \tau_1) - \ell_{x}(q, \tau_2)\right]\right|
    \le \int \mathcal{P}(x) \left|\ell_{x}(q, \tau_1) - \ell_{x}(q, \tau_2)\right|  dx
    \le  L \left|\tau_1 - \tau_2\right| \int \mathcal{P}(x) dx
    =  L \left|\tau_1 - \tau_2\right|,
\end{gather*}
completing the proof as desired.
\end{proof}

\subsection{Predictive Efficiency of \textsc{CANVI}}\label{section:empirical_iterate_thm}

The proof emerges through a reduction of \textsc{CANVI} to a special case of the ``Validity First Conformal Prediction for the Smallest Prediction Set'' (VFCP) algorithm presented in the \citep{yang2021finite}. The VFCP algorithm is provided for convenience in Algorithm \ref{alg:valid_first}. Note that we made the appropriate replacements in notations to match those presented in the main body of our paper for clarity.

To further underscore the parallel, we present an immaterially modified version of \textsc{VFCP}, where it is assumed the data are split \textit{prior} to the algorithm execution into disjoint sets $\mathcal{D},\mathcal{D}_{\mathcal{C}}$, and $\mathcal{D}_{\mathcal{R}}$. We assume input prediction methods $\mathcal{A}^{(1)}, ..., \mathcal{A}^{(T)}$ have been trained on $\mathcal{D}$. Finally, the presentation in \citep{yang2021finite} allows for generic definitions of the measure function of $\mathcal{C}(x)$, referred to as ``Width'' therein. We present it with the particular Lebesgue measure function as presented in the main body to avoid confusion. These three modifications in the algorithm presentation have no manifest effects in the proof of \citep{yang2021finite}, meaning all results as presented there apply with the appropriate choices of inputs in Algorithm \ref{alg:valid_first}. 

We similarly consider an immaterially modified form of \textsc{CANVI}, that is, one in which $\mathcal{D}_{\mathcal{C}}$ and $\mathcal{D}_{\mathcal{R}}$ are pre-generated in precisely the fashion outlined Algorithm \ref{alg:canvi} and provided as input. Thus, inputs to \textsc{CANVI} parallel those of \textsc{VFCP}, namely $\{q^{(t)}(\Theta\mid X)\}_{t=1}^T,1-\alpha,D_{\mathcal{C}},$ and $D_{\mathcal{R}}$.

Intuitively, the proof follows from the fact that \textsc{CANVI} is a special case of Algorithm \ref{alg:valid_first}, where the particular structural assumptions we make imply the corresponding assumptions as stated in the theorem from \citep{yang2021finite}. Algorithm \ref{alg:valid_first} is presented using the so-called ``nested set formulation'' of conformal prediction. Briefly, the nested sets formulation starts by requiring the user to specify a \textit{set-valued} function $F_r$ instead of a score function. Sets must be nested over increasing $r$. Despite being seemingly more expressive, the equivalence of the nested set and split score conformal methods was demonstrated in \citep{gupta2022nested}. In particular, for a chosen score function $s(x, \theta)$, the corresponding nested set function would be $F^{(t)}_{\tau} = \{(x, \theta) \mid s^{(t)}(x, \theta) \le \tau\}$, where $s^{(t)}$ denotes the score function as defined with $\mathcal{A}^{(t)}$. The presentation of Algorithm \ref{alg:valid_first} using the nested set formulation was, thus, a means to allow for the employment of relevant proof strategies from classical learning theory. We now proceed through the formal equivalence of \textsc{CANVI} and a special case of \textsc{VFCP}.

\begin{algorithm}
  \caption{\label{alg:valid_first} Validity First Conformal Prediction (VFCP) \citep{yang2021finite}}
  \begin{algorithmic}[1]
    \STATE{\textbf{Procedure:} \textsc{VFCP}}
    \STATE{\textbf{Inputs:} Predictors $\{\mathcal{A}^{(t)}\}_{t=1}^{T}$, Target coverage $1-\alpha$, Calibration set $D_{\mathcal{C}}$, Recalibration set $D_{\mathcal{R}}$, Score $s(x, \theta)$}
    \STATE{Using $\mathcal{A}^{(t)}$, construct an increasing
    (nested) sequence of sets $\{F^{(t)}_{\tau}\}_{\tau\in\mathcal{T}}$, where $\mathcal{T}\subset \mathbb{R}$ and 
    $$ F^{(t)}_{\tau} = \{(x, \theta) \mid s^{(t)}(x, \theta) \le \tau\}, $$
    where $s^{(t)}$ denotes the score function as defined with $\mathcal{A}^{(t)}$}
    \STATE{Compute conformal prediction set $\mathcal{C}^{(t)}$ based on $\{F^{(t)}_{\tau}\}_{\tau\in\mathcal{T}}$. Specifically, for each $(x_i, \theta_i) \in \mathcal{D}_{\mathcal{C}}$ and $t \in \{1,2,...,T\}$, denote its corresponding score as
    $$ s^{(t)}(x_i, \theta_i) := \inf_{\tau \in\mathcal{T}}\{(x_i, \theta_i) \in F^{(t)}_{\tau} \}$$}
    \STATE{Compute the corresponding conformal prediction set as 
    $$\mathcal{C}_{\mathcal{C}}^{(t)} := \{(x, \theta) : s^{(t)}(x, \theta) \le \widehat{q}_{\mathcal{C}}^{(t)}(\alpha)\} ,$$ 
    where $\widehat{q}_{\mathcal{C}}^{(t)}(\alpha)$ is
    the $\lceil(|\mathcal{D}_{\mathcal{C}}|+1)(1 - \alpha)\rceil$-th largest element of $\{s^{(t)}(x_i, \theta_i)\}_{i\in \mathcal{D}_{\mathcal{C}}}$}
    \STATE{Let $\mathcal{C}_{\mathcal{C}}^{(t)}(x) := \{\theta : (x, \theta) \in \mathcal{C}_{\mathcal{C}}^{(t)}\}$. Set
    $$t^{*} := \argmin_{1\le t\le T} \mathbb{E}_{X}[\mathcal{L}(\mathcal{C}_{\mathcal{C}}^{(t)}(X))] $$}
    \STATE{For each $(x_i, \theta_i) \in \mathcal{D}_{\mathcal{R}}$, define the conformal score
    $$ s^{(*)}(x_i, \theta_i) := \inf_{\tau \in\mathcal{T}}\{(x_i, \theta_i) \in F^{(t^{*})}_{\tau} \}$$}
    \STATE{Compute the corresponding conformal prediction set as
    $$\mathcal{C}_{\mathcal{R}}^{(*)} := \{(x, \theta) : s^{(*)}(x, \theta) \le \widehat{q}_{\mathcal{R}}^{(*)}(\alpha)\} ,$$
    where $\widehat{q}_{\mathcal{R}}^{(*)}(\alpha) := \lceil( |\mathcal{D}_{\mathcal{R}}|+1)(1 - \alpha)\rceil$-th largest element of $\{s^{(*)}(x_i, \theta_i)\}_{i\in\mathcal{D}_{\mathcal{R}}}$}
    \STATE {Return the prediction set $\mathcal{C}_{\mathcal{R}}^{(*)}$}
    \end{algorithmic}
\end{algorithm}

We first state for reference the original theorem for VFCP, which we leverage in the proof of theorem.

\begin{theorem}\label{theorem:vfcp}
    Suppose Assumption \ref{assump:opt_technical_2} holds. Let $\alpha\in(0,1)$, $\mathcal{D_C}$ and $\mathcal{D_R}$ be drawn i.i.d. from $\mathcal{P}(X, \Theta)$, and
    \begin{gather*}
        \mathcal{C}_{\mathcal{R}}^{(*)} =
        \mathrm{VFCP}\left(
        \mathcal{A}^{(t)},1-\alpha,D_{\mathcal{C}},D_{\mathcal{T}},s(x, \theta)
    \right)
    \end{gather*}    
    If, for
        $r\ge\max\{\sqrt{\log(4T/\delta)/2N_{\mathcal{C}}}, 2 / N_{\mathcal{C}}\}$
    and $\delta\in[0,1]$, Assumption \ref{assump:opt_technical_1} holds, then with probability at least $(1-\delta)$,
    \begin{equation}
    \begin{gathered}
        \mathbb{E}_{X}[\mathcal{L}(\mathcal{C}_{\mathcal{R}}^{(*)}(X))] \le
        \min_{1\le t\le T} \mathbb{E}_{X}[\mathcal{L}(\mathcal{C}_{\mathcal{R}}^{(t)}(X))] 
        + 3 L_W L_{[T]} \left[\left(\frac{\log(4T/\delta)}{N_{\mathcal{C}}}\right)^{\gamma/2} + \left(\frac{2}{N_{\mathcal{C}}}\right)^{\gamma}\right],        
    \end{gathered}
    \end{equation}
    where $\gamma$, $L_W$, and $L_{[T]} = \max_{1\le t\le T} L_t$ are constants defined in Assumptions \ref{assump:opt_technical_1} and \ref{assump:opt_technical_2}  \citep{yang2021finite}.
\end{theorem}

We now present the proof of our theorem. The proof proceeds in two steps. We first demonstrate the \textsc{CANVI} and VFCP algorithms are equivalent if we assume access to the exact $\ell(q, \tau)$, which, coupled with the assumptions imposed on $q$, allows us to directly leverage the results of Theorem \ref{theorem:vfcp}. We then demonstrate, under Assumption \ref{assump:opt_technical_3}, we recover the desired bound even if $t^{*}$ is chosen with $\widehat{\ell}(q, \tau)$.

\begin{theorem}
    Suppose for any $x\in\mathcal{X}$ and $t=1,...,T$, $q^{(t)}(\theta\mid x)\in\mathcal{C}^{3}(\mathbb{R}^{n})$ is bounded above and for $\theta\neq0$, $\mathcal{L}(\{\theta : \nabla_{\theta} q^{(t)}(\theta|x)\}) = 0$. Further assume $P(X,\Theta)$ is bounded above. 
    Let $\alpha\in(0,1)$ and
    \begin{gather*}
        q^{(*)}(\Theta\mid X),\widehat{q}_{\mathcal{R}}^{(*)}(\alpha) = \\
        \mathrm{CANVI}\left(
        \{q^{(t)}(\Theta\mid X)\}_{t=1}^T,\mathcal{P}(X, \Theta),1-\alpha,N_{\mathcal{C}},N_{\mathcal{T}}
    \right)
    \end{gather*}    
    If, for
        $r\ge\max\{\sqrt{\log(4T/\delta)/2N_{\mathcal{C}}}, 2 / N_{\mathcal{C}}\}$
    and $\delta\in[0,1]$, Assumption \ref{assump:opt_technical_1} holds
    and for $\Delta, \epsilon > 0$ Assumption \ref{assump:opt_technical_3} holds, then with probability at least $(1-\epsilon)(1-\delta)$,
    \begin{equation}
    \begin{gathered}
        \ell(q^{(*)},\widehat{q}_{\mathcal{R}}^{(*)}(\alpha)) \le
        \min_{1\le t\le T} \ell(q^{(t)},\widehat{q}_{\mathcal{R}}^{(t)}(\alpha)) + \Delta 
        + 3 L_W L_{[T]} \left[\left(\frac{\log(4T/\delta)}{N_{\mathcal{C}}}\right)^{\gamma/2} + \left(\frac{2}{N_{\mathcal{C}}}\right)^{\gamma}\right],        
    \end{gathered}
    \end{equation}
    where $\gamma$, $L_W$, and $L_{[T]} = \max_{1\le t\le T} L_t$ are constants defined in Assumptions \ref{assump:opt_technical_1} and \ref{assump:opt_technical_2}.
\end{theorem}

\begin{proof}
    Let
        $\mathcal{C}_{\mathcal{R}}^{(*)} =
        \mathrm{VFCP}\left(
        \{q^{(t)}(\Theta\mid X)\}_{t=1}^T,1-\alpha,D_{\mathcal{C}},D_{\mathcal{T}}, 1 / q(\theta\mid x)
    \right)$.
    We first wish to demonstrate $\mathcal{C}_{\mathcal{C},\mathcal{R}}^{(t),\textsc{VFCP}}(x) = \mathcal{C}_{\mathcal{C},\mathcal{R}}^{(t),\textsc{CANVI}}(x)$ for any fixed $x$, where we use the $\mathcal{C},\mathcal{R}$ condensed notation to mean this equality holds both under $\mathcal{D}_{\mathcal{C}}$ and $\mathcal{D}_{\mathcal{R}}$. This equivalence can be shown in demonstrating the equivalence in corresponding scores, as the resulting empirical score distributions and hence quantiles over $\mathcal{D}_\mathcal{C}$ and $\mathcal{D}_\mathcal{R}$
    and finally future prediction regions follow to then be equivalent by the algorithm structure. 

    

    This equivalence follows as a straightforward instance of the equivalence of the set-valued and standard conformal prediction frameworks. In particular, we recover the original score formulation, as:
    \begin{gather*}
        s^{(t)}(x_i, \theta_i) 
        := \inf_{\tau \in\mathcal{T}}\{(x_i, \theta_i) \in F^{(t)}_{\tau} \}
        = \inf_{\tau \in\mathcal{T}}\{(x_i, \theta_i) \in \{(x, \theta) \mid 1 / q^{(t)}(\theta\mid x) \le \tau\} \}
        =  1 / q^{(t)}(\theta_i\mid x_i).
    \end{gather*}
    The bound under access to the exact $\ell(q, \tau)$ then follows from Theorem \ref{theorem:vfcp} under demonstration of the appropriate assumptions. Assumption \ref{assump:opt_technical_1} holds by assumption. Assumption \ref{assump:opt_technical_2} holds for any $\vartheta^{(t)} := \{\theta : \nabla_{\theta} q^{(t)}(\theta|x)\}$ by Corollary \ref{corollary:smoothness}. In the application of Assumption \ref{assump:opt_technical_2} for the proof of Theorem \ref{theorem:vfcp}, it suffices for $\mathcal{F}_{t}^{-1}(1 - \alpha)\in\vartheta^{(t)}$ and $\mathcal{F}_{t}^{-1}(1 - \alpha + 1/(N_{\mathcal{C}} + 1))\in\vartheta^{(t)}$. Since $\mathcal{L}(\{\theta : \nabla_{\theta} q^{(t)}(\theta|x)\}) = 0$ and $\mathcal{P}(X, \Theta)$ is bounded above, this means $\mathcal{P}_{X, \Theta}(\mathcal{F}_{t}^{-1}(1 - \alpha) \in\vartheta^{(t)}) = 1$ and $\mathcal{P}_{X, \Theta}(\mathcal{F}_{t}^{-1}(1 - \alpha + 1/(N_{\mathcal{C}} + 1))\in\vartheta^{(t)}) = 1$. Thus, if $t^{*}$ is chosen in \textsc{CANVI} using $\ell(q, \tau)$, by Theorem \ref{theorem:vfcp}, with probability $1(1-\delta) = 1-\delta$,
    \begin{gather*}
        \ell(q^{(*)},\widehat{q}_{\mathcal{R}}^{(*)}(\alpha)) \le
        \min_{1\le t\le T} \ell(q^{(t)},\widehat{q}_{\mathcal{R}}^{(t)}(\alpha)) 
        + 3 L_W L_{[T]} \left[\left(\frac{\log(4T/\delta)}{N_{\mathcal{C}}}\right)^{\gamma/2} + \left(\frac{2}{N_{\mathcal{C}}}\right)^{\gamma}\right].
    \end{gather*}
    The extension of this to the case of interest, where $t^{*}$ is chosen using $\widehat{\ell}(q, \tau)$, is now a straightforward application of Assumption \ref{assump:opt_technical_3}, from which we have that $\exists$ $\Delta, \epsilon > 0$, such that with probability at least $1-\epsilon$
    \begin{equation}
        \left|\ell(q^{(\widehat{t}^{*})}, \widehat{q}_\mathcal{R}^{(\widehat{t}^{*})}(\alpha)) - \ell(q^{(t^{*})}, \widehat{q}_\mathcal{R}^{(t^{*})}(\alpha))\right| < \Delta
        \implies \ell(q^{(\widehat{t}^{*})}, \widehat{q}_\mathcal{R}^{(\widehat{t}^{*})}(\alpha)) < \ell(q^{(t^{*})}, \widehat{q}_\mathcal{R}^{(t^{*})}(\alpha)) + \Delta.
    \end{equation}    
    Switching back to denoting $\ell(q^{(*)},\widehat{q}_{\mathcal{R}}^{(*)}(\alpha)) := \ell(q^{(\widehat{t}^{*})}, \widehat{q}_\mathcal{R}^{(\widehat{t}^{*})}(\alpha))$, this implies that, with probability $(1-\epsilon)(1-\delta)$,
    \begin{gather*}
        \ell(q^{(*)},\widehat{q}_{\mathcal{R}}^{(*)}(\alpha)) \le
        \ell(q^{(t^{*})}, \widehat{q}_\mathcal{R}^{(t^{*})}(\alpha)) + \Delta
         \le
        \min_{1\le t\le T} \ell(q^{(t)},\widehat{q}_{\mathcal{R}}^{(t)}(\alpha)) + \Delta
        + 3 L_W L_{[T]} \left[\left(\frac{\log(4T/\delta)}{N_{\mathcal{C}}}\right)^{\gamma/2} + \left(\frac{2}{N_{\mathcal{C}}}\right)^{\gamma}\right],
    \end{gather*}
    completing the proof as desired.
\end{proof}

\section{Gaussian Hölder Continuity}\label{section:informative_thm_detailed}

\begin{theorem}
Let $\Theta$ and $X$ be zero-mean unit-variance Gaussian random variables with correlation $\rho$. Let $q^{(t)}(\theta|x)=\mathcal{N}(\theta;t x, 1 - \rho^2)$. Let $\kappa := t^2 - 2 t \rho + 1$ and $r > 0$. Then $F_t^{-1}(z)$, is $1$-Hölder continuous on \upshape[$ 1-\alpha,1-\alpha + r$\upshape] \itshape with Hölder constant
\begin{equation}
    \frac{\kappa \Phi^{-1}(\frac{1-\alpha}{2}) \sqrt{
    \exp\left(\frac{\kappa}{1-\rho^2} \Phi^{-1}(\frac{1-\alpha}{2})^2 -\frac{(1-\alpha)^2}{2}\right)
    }
    }{\sqrt{(1-\rho^2) / 2}}
\end{equation}
\end{theorem}

\begin{proof}
Notice that in the bivariate Gaussian case, we have closed forms for the following:
\begin{gather*}
    \Theta\mid x\sim\mathcal{N}(\rho x, 1-\rho^2)\qquad
    X\mid \theta\sim\mathcal{N}(\rho \theta, 1-\rho^2)\\
    \Theta \sim\mathcal{N}(0, 1) \qquad
    X \sim\mathcal{N}(0, 1).
\end{gather*}
We wish to find the distribution of $s(X, \Theta) = 1 / q(\Theta\mid X)$ jointly over $X, \Theta$ to find $F_t^{-1}(z)$ explicitly. The CDF of this score can be computed as follows:
\begin{gather*}
    \mathcal{P}\left(1 / q_t(\Theta\mid X) \le q\right)
    = \mathcal{P}\left(\sqrt{2\pi(1-\rho^2)} e^{\frac{(t X-\Theta)^2}{2(1-\rho^2)}} \le q\right) \\
    = \mathcal{P}\left(R^2 \le 2(1-\rho^2) \log\left(\sqrt{\frac{q^2}{2\pi(1-\rho^2)}}\right)\right)
    = \mathcal{P}\left(R \le \sqrt{(1-\rho^2) \log\left(\frac{q^2}{2\pi(1-\rho^2)}\right)}\right),
\end{gather*}
where $R := |t X-\Theta|$. From the above calculation, $F_t^{-1}(z)$ must satisfy:
$$ 
\mathcal{P}\left(R \le \sqrt{(1-\rho^2) \log\left(\frac{F_t^{-1}(z)^2}{2\pi(1-\rho^2)}\right)}\right) = z.
$$
Notice now that, since $(X, \Theta)$ are bivariate Gaussian:
\begin{align*}
    t X-\Theta \sim\mathcal{N}(0, t^2 + 1 - 2t \rho)
    \implies R\sim\text{HalfNormal}(t^2 + 1 - 2t \rho).
\end{align*}
Therefore, the $z$ quantile of $R$ is $\sqrt{t^2 + 1 - 2t \rho} \Phi^{-1}\left(1-\frac{\alpha}{2}\right)$. Solving for $F_t^{-1}(z)$ in this quantile produces the final threshold:
\begin{align*}
    \sqrt{(1-\rho^2) \log\left(\frac{F_t^{-1}(z)^2}{2\pi(1-\rho^2)}\right)} = \sqrt{t^2 + 1 - 2t \rho} \Phi^{-1}\left(1-\frac{\alpha}{2}\right) \\
    \implies
    \log\left(\frac{F_t^{-1}(z)^2}{2\pi(1-\rho^2)}\right) = \frac{t^2 + 1 - 2t \rho}{1-\rho^2} \left(\Phi^{-1}\left(1-\frac{\alpha}{2}\right)\right)^2 \\
    \implies
    F_t^{-1}(z) = \sqrt{2\pi(1-\rho^2) \exp\left(\frac{t^2 + 1 - 2t \rho}{1-\rho^2} \left(\Phi^{-1}\left(1-\frac{\alpha}{2}\right)\right)^2\right)}.
\end{align*}

The Hölder constant follows from bounding the derivative of $F^{-1}_t(z)$, namely
$$
\frac{\kappa \Phi^{-1}(\frac{z}{2}) \sqrt{
    \exp\left(\frac{\kappa}{1-\rho^2} \Phi^{-1}(\frac{z}{2})^2 -\frac{(z)^2}{2}\right)
    }
    }{\sqrt{(1-\rho^2) / 2}}
$$
This expression is monotonically decreasing in $z$ and hence maximized for $z = 1-\alpha$ in the interval, giving the desired expression.

\end{proof}

\section{Simulation-Based Inference Benchmarks}\label{section:benchmark}
The benchmark tasks are a subset of those provided by \citep{lueckmann2021benchmarking}. For convenience, we provide brief descriptions of the tasks curated by this library; however, a more comprehensive description of these tasks can be found in their manuscript.

\subsection{Gaussian Linear}
10-dimensional Gaussian model with a Gaussian prior:

\begin{gather*}
    \text{\textbf{Prior}: } \mathcal{N}(0, 0.1 \odot I)\\
    \text{\textbf{Simulator}: } x\mid w \sim \mathcal{N}(x\mid w, 0.1 \odot I)
\end{gather*}

\subsection{Gaussian Linear Uniform}
10-dimensional Gaussian model with a uniform prior:

\begin{gather*}
    \text{\textbf{Prior}: } \mathcal{U}(-1, 1)\\
    \text{\textbf{Simulator}: } x\mid w \sim \mathcal{N}(x\mid w, 0.1 \odot I)
\end{gather*}



\subsection{SLCP with Distractors}
Simple Likelihood Complex Posterior (SLCP) with Distractors has uninformative dimensions in the observation over the standard SLCP task:

\begin{gather*}
    \text{\textbf{Prior}: } \mathcal{U}(-3, 3)  \\
    \text{\textbf{Simulator}: } x\mid w = p(y)
    \text{ where } p \text{ reorders } \\
    y \text{ with a fixed random order } \\
    y_{[1:8]} \sim \mathcal{N}\left(
        \begin{bmatrix}
        w_1 \\ 
        w_2
        \end{bmatrix}, 
    \begin{bmatrix}
        w_3^4 & w_3^2 w_4^2 \tanh(w_5) \\ 
        w_3^2 w_4^2 \tanh(w_5) & w_4^4
    \end{bmatrix}\right),\\
    y_{9:100} \sim \frac{1}{20} \sum_{i=1}^{20} t_2(\mu^i, \Sigma^i), 
    \mu^i\sim\mathcal{N}(0, 15^2 I),  \\
    \Sigma^i_{j,k}\sim\mathcal{N}(0, 9),
    \Sigma^i_{j,j} = 3 e^{a}, a\sim\mathcal{N}(0, 1),
\end{gather*}



\subsection{Bernoulli GLM Raw}
10-parameter GLM with Bernoulli observations and Gaussian prior. Observations are not sufficient statistics, unlike the standard ``Bernoulli GLM'' task:

\begin{gather*}
    \text{\textbf{Prior}: } 
    \beta\sim\mathcal{N}(0, 2), f\sim\mathcal{N}(0, (F^T F)^{-1}) \\
    \qquad F_{i,i-2} = 1, F_{i,i-1} = -2 \\
    F_{i,i} = 1 + \sqrt{\frac{i-1}{9}}, F_{i, j} = 0; i\le j \\
    \text{\textbf{Simulator}: } x^{(i)} \mid w \sim \text{Bern}(\eta(v^{(i)}_T f+ \beta)), \\ \eta(\odot) = \exp(\odot) / (1 + \exp(\odot)) 
\end{gather*}

\subsection{Gaussian Mixture}
A mixture of two Gaussians, with one having a much broader covariance structure:

\begin{gather*}
    \text{\textbf{Prior}: } 
    \beta\sim\mathcal{U}(-10, 10) \\
    \text{\textbf{Simulator}: } x \mid w \sim 0.5 \mathcal{N}(x\mid w, I) + 0.5 \mathcal{N}(x\mid w, .01 I)
\end{gather*}

\subsection{Two Moons}
Task with a posterior that has both global (bimodal) and local (crescent-shaped) structure:

\begin{gather*}
    \text{\textbf{Prior}: } 
    \beta\sim\mathcal{U}(-1, 1) \\
    \text{\textbf{Simulator}: } x \mid w =  \\
    \begin{bmatrix}
        r \cos(\alpha) + 0.25 \\ 
        r \sin(\alpha)
    \end{bmatrix} +
    \begin{bmatrix}
        -|w_1 + w_2| / \sqrt{2} \\ 
        (-w_1 + w_2) / \sqrt{2}
    \end{bmatrix}\\
    \qquad \alpha\sim \mathcal{U}(-\pi/2,\pi/2), r\sim\mathcal{N}(0.1,0.01^2)
\end{gather*}

\subsection{SIR}
Epidemiology model with $S$ (susceptible), $I$ (infected), and $R$ (recovered). A contact rate $\beta$ and mean recovery rate of $\gamma$ are used as follows:

\begin{gather*}
    \text{\textbf{Prior}: } 
    \beta\sim\text{LogNormal}(\log(0.4),0.5), \\
    \gamma\sim\text{LogNormal}(\log(1/8),0.2) \\
    \text{\textbf{Simulator}: } x = (x^{(i)})_{i=1}^{10}; x^{(i)} \mid w\sim\text{Bin}(1000, \frac{I}{N}),  \\
    \text{ where } I \text{ is simulated from: } 
    \\
    \frac{dS}{dt} = -\beta \frac{SI}{N}, \qquad
    \frac{dI}{dt} = \beta \frac{SI}{N} - \gamma I, \qquad
    \frac{dR}{dt} = \gamma I
\end{gather*}

\subsection{Lotka-Volterra}
An ecological model commonly used in describing dynamics of competing species. $w$ parameterizes this interaction as $w = (\alpha, \beta, \gamma, \delta)$:

\begin{gather*}
    \text{\textbf{Prior}: } 
    \alpha\sim\text{LogNormal}(-.125,0.5) \\
    \beta\sim\text{LogNormal}(-3,0.5),
    \gamma\sim\text{LogNormal}(-.125,0.5) \\
    \delta\sim\text{LogNormal}(-3,0.5) \\
    \text{\textbf{Simulator}: } x = (x^{(i)})_{i=1}^{10},  \\
    x_{1,i} \mid w\sim\text{LogNormal}(\log(X), 0.1),  \\
    x_{2,i} \mid w\sim\text{LogNormal}(\log(Y), 0.1) \\ 
    \text{ where } X,Y \text{ is simulated from: }  \\
    \frac{dX}{dt} = \alpha X - \beta X Y, \qquad
    \frac{dY}{dt} = - \gamma Y + \delta XY
\end{gather*}

\subsection{ARCH}\label{section:arch_details}
The two-dimensional parameter $\theta = (\theta_1, \theta_2)$ includes both an autoregressive component ($\theta_1$) and a component controlling the level of conditional noise ($\theta_2$). Given a full realization of the time series $y_{1:T}$, we aim to amortize inference over $\theta$. Priors are taken to be $\theta_1 \sim \textrm{Unif}(-1, 1)$ and $ \theta_2 \sim \textrm{Unif}(0, 1)$. One important change from the model of \citep{owen2022lfire} is that we fix $e^{(0)}=0$ rather than drawing this quantity from a standard Gaussian.

The Adam optimizer was used with a learning rate of 0.0001 for 25,000 training steps for each of the three methods: IWBO ($K=10$), ELBO, FAVI. Due to constraints arising from the uniform priors on the parameters, the IWBO and ELBO implementations rely on logit transformations of the latent random variables $\theta$ to avoid zero-density regions that result in undefined gradients. The encoder network is trained to learn distributions on the unconstrained space of the transformed random variable $\theta'$, and visualizations are produced by performing the inverse transformation. While FAVI avoids these issues because it is likelihood-free, for an apples-to-apples comparison we also implement FAVI on the unconstrained latent space as well. 

\section{Training Details}\label{section:exp_details}
All encoders were implemented in PyTorch \citep{paszke2019pytorch} with a Neural Spline Flow architecture. The NSF was built using code from \citep{nflows}. Specific architecture hyperparameter choices were taken to be the defaults from \citep{nflows} and are available in the code. Optimization was done using Adam \citep{kingma2014adam} with a learning rate of $10^{-3}$ over 5,000 training steps. Minibatches were drawn from the corresponding prior $\mathcal{P}(\Theta)$ and simulator $\mathcal{P}(X\mid\Theta)$ as specified per task in the preceding section. Training these models required between 10 minutes and two hours using an Nvidia RTX 2080 Ti GPUs for each of the SBI tasks.

\section{Prediction Regions}\label{section:credible_regions}
Credible regions for $q_\varphi(\Theta\mid x)$ (for a single $x\sim\mathcal{P}(X)$) on a subset of the SBI benchmark tasks are plotted below over varying degrees of training, as indicated in the figures. As expected, the efficiency of the prediction regions improves over training across all tasks, resulting in \textit{smaller} regions for each target coverage $1-\alpha$.

\subsection{Gaussian Linear}
\begin{figure}[H]
\includegraphics[scale=0.16]{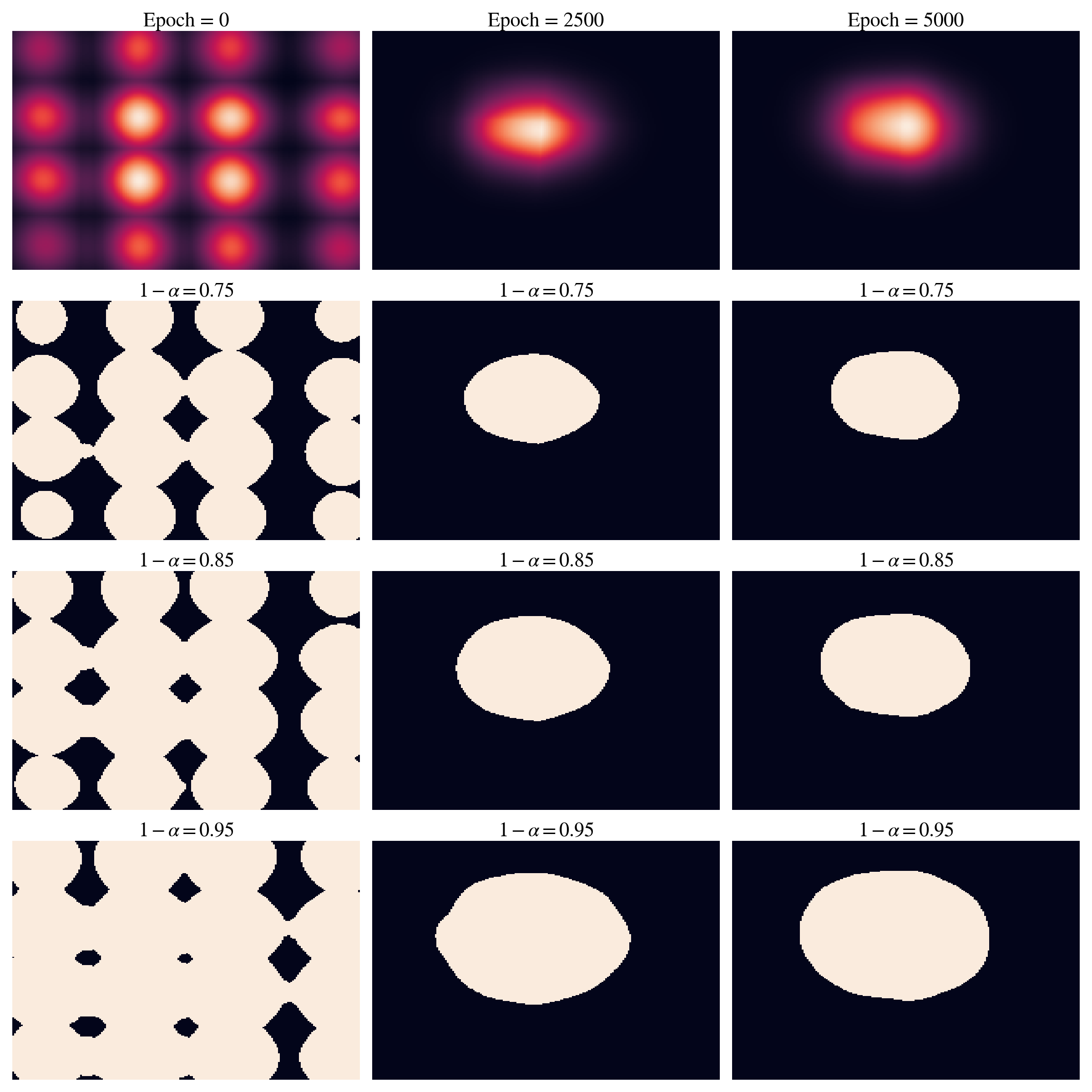}
\centering
\end{figure}

\subsection{Gaussian Linear Uniform}
\begin{figure}[H]
\includegraphics[scale=0.16]{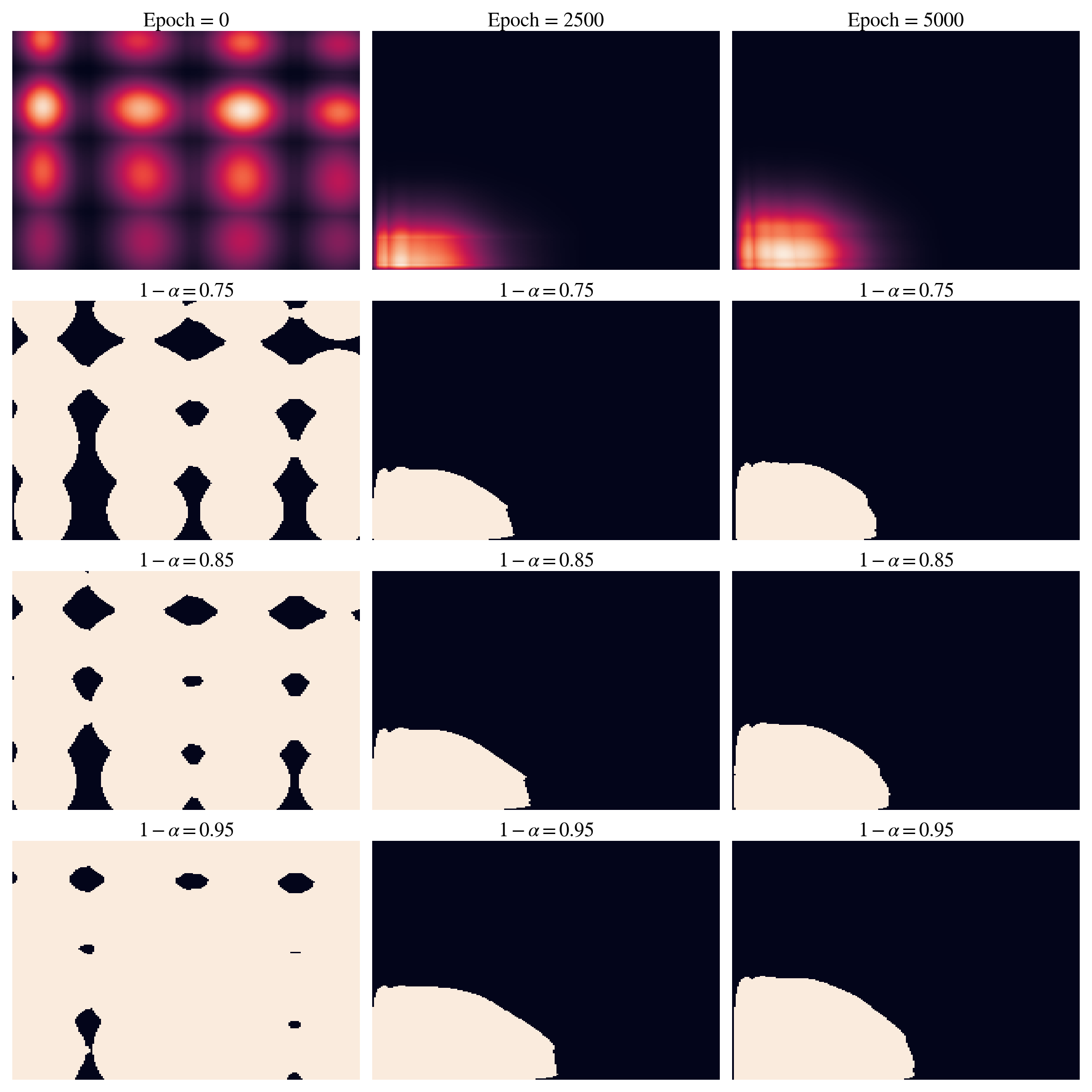}
\centering
\end{figure}

\subsection{SLCP}
\begin{figure}[H]
\includegraphics[scale=0.16]{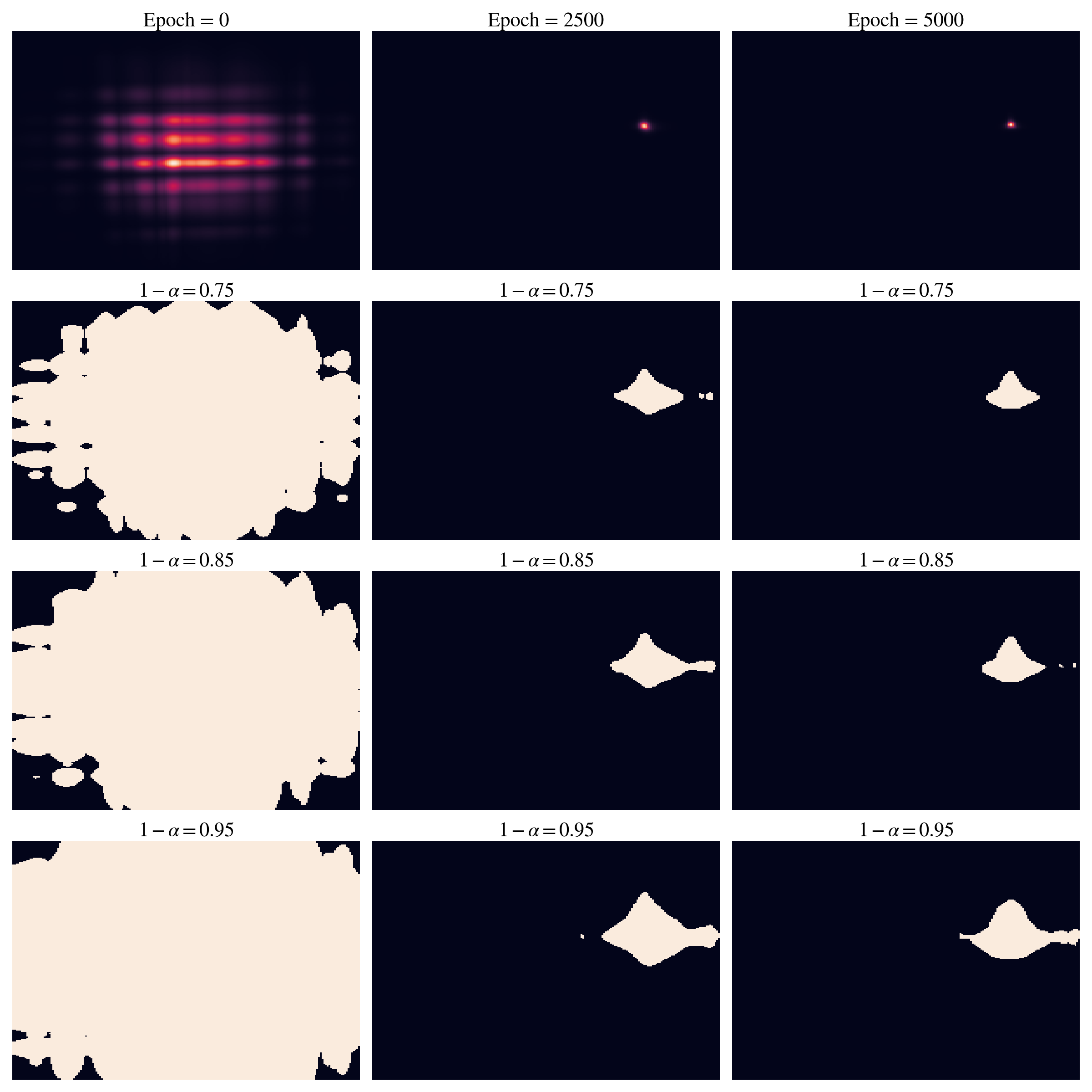}
\centering
\end{figure}

\subsection{Bernoulli GLM Raw}
\begin{figure}[H]
\includegraphics[scale=0.16]{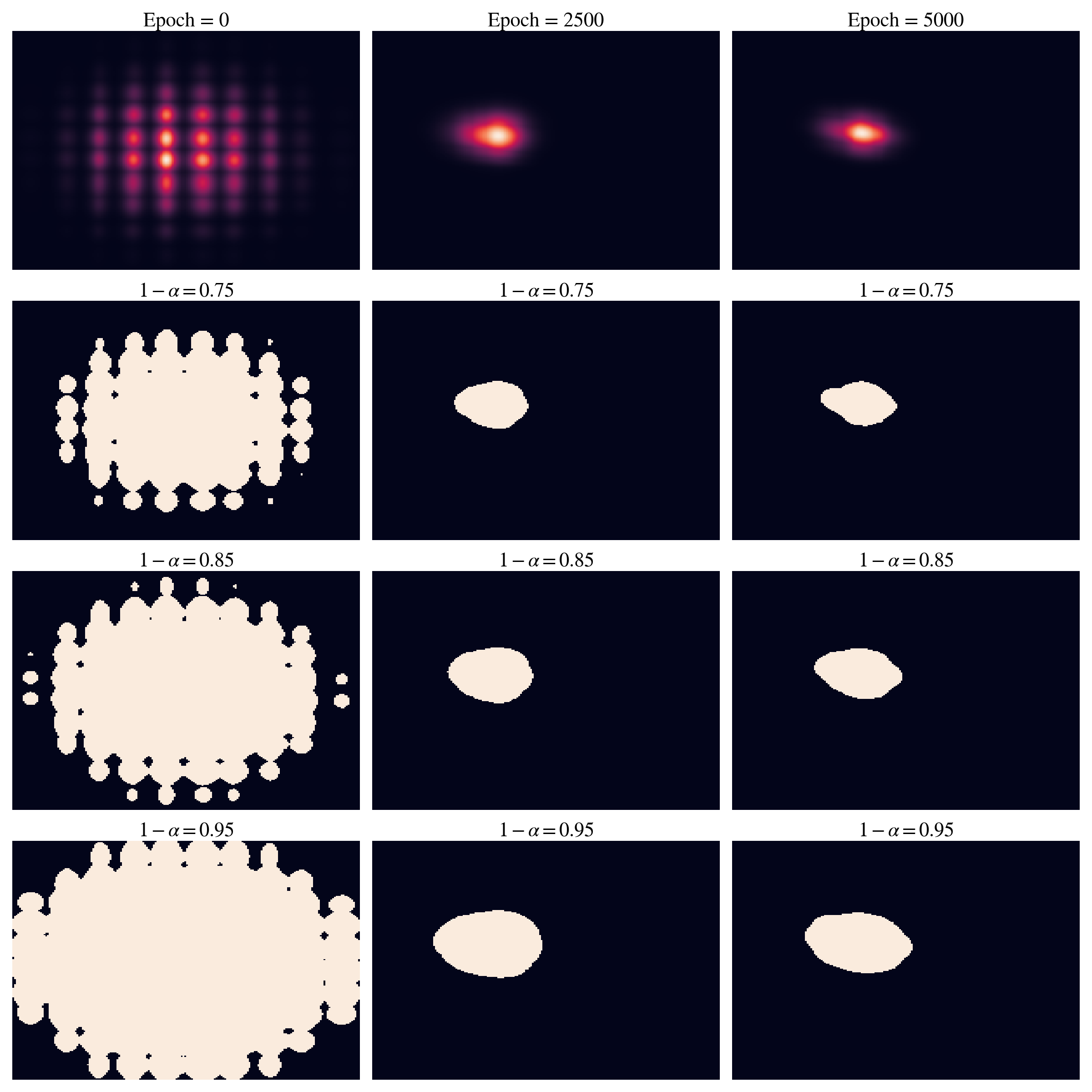}
\centering
\end{figure}

\subsection{Gaussian Mixture}
\begin{figure}[H]
\includegraphics[scale=0.16]{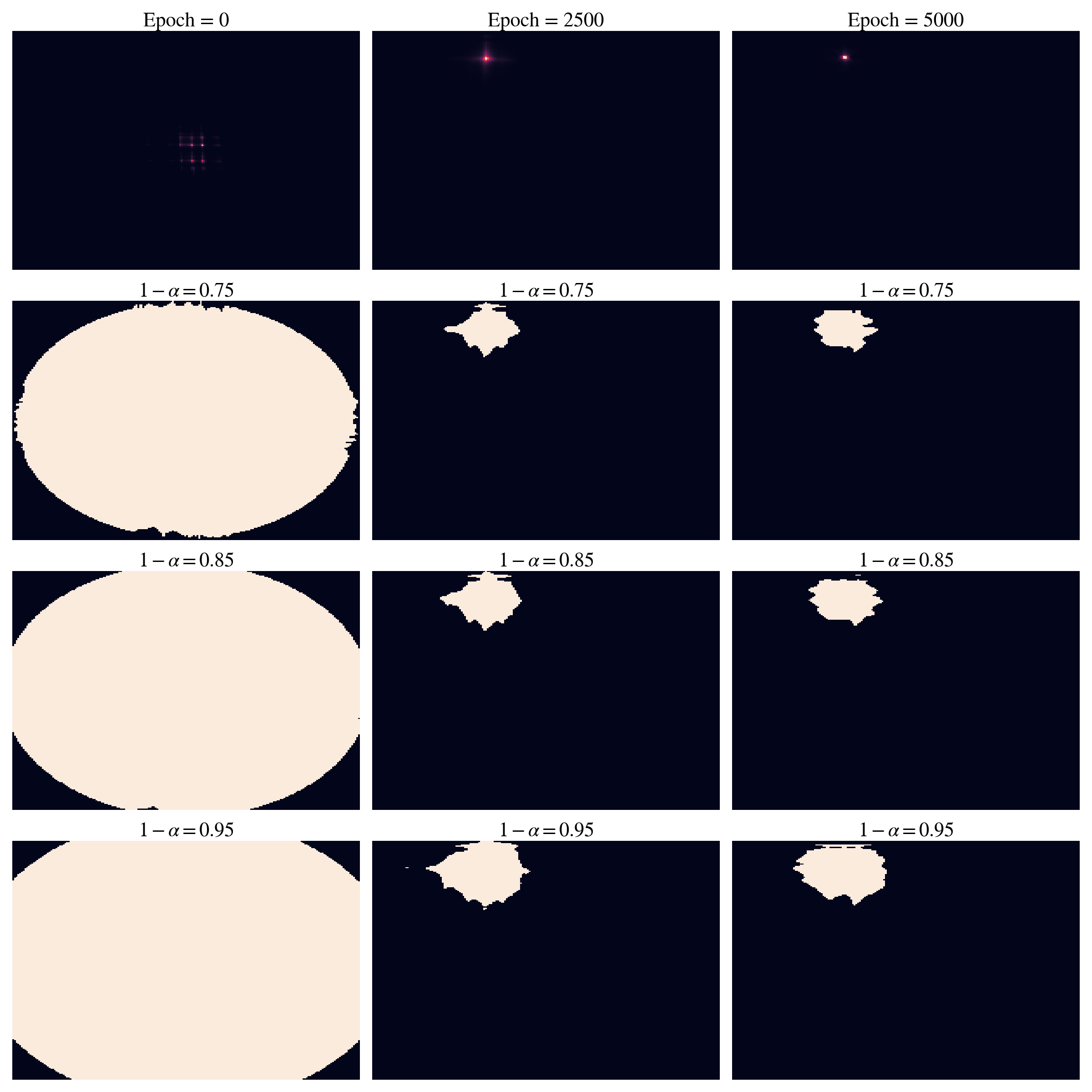}
\centering
\end{figure}

\subsection{Two Moons}
\begin{figure}[H]
\includegraphics[scale=0.16]{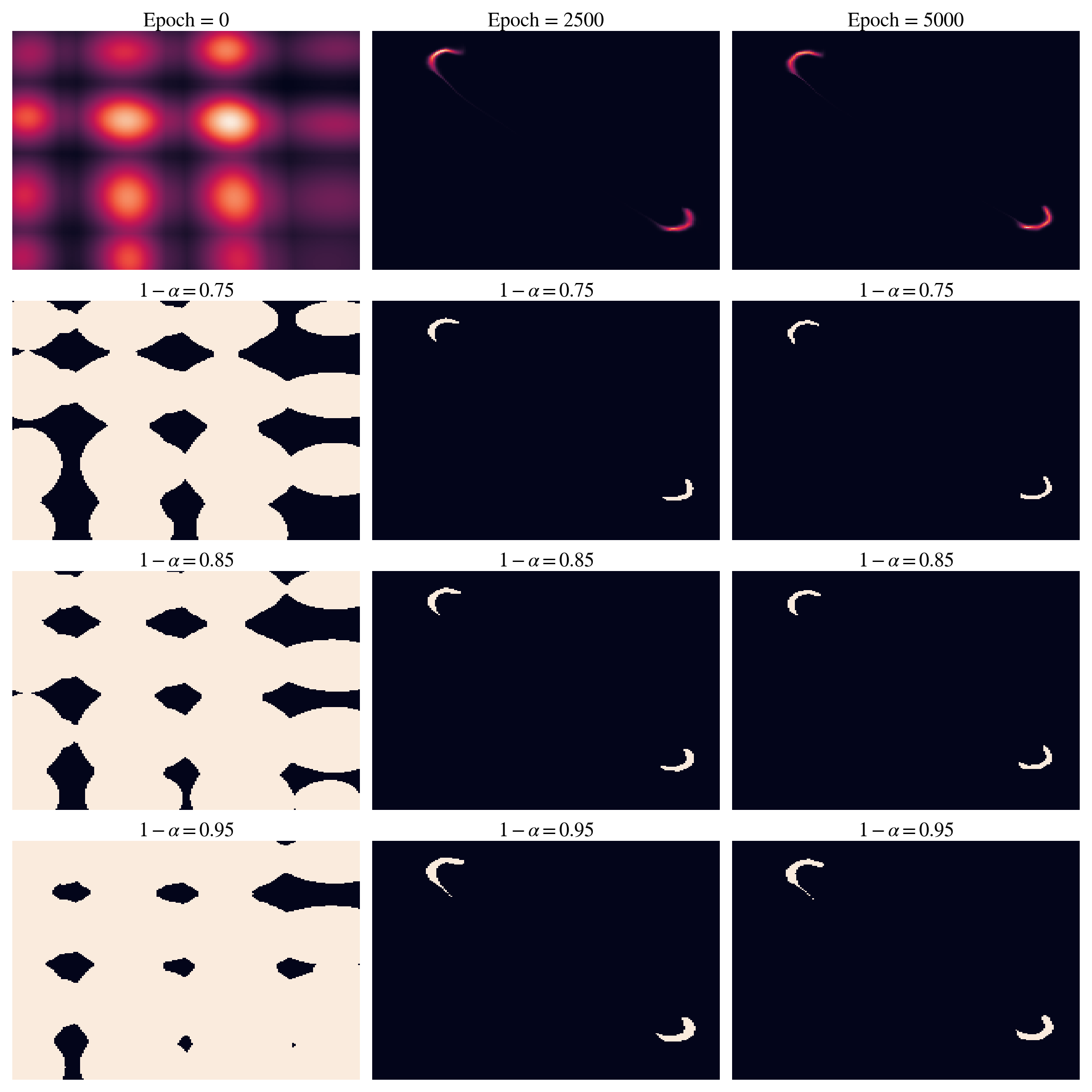}
\centering
\end{figure}

\section{Posteriors}\label{section:posteriors}
We provide visualizations of approximate and reference posteriors (produced with MCMC from \citep{lueckmann2021benchmarking}) to justify the overdispersion claims made on the variational approximation procedure.

\subsection{Gaussian Linear}
\begin{figure}[H]
\includegraphics[scale=0.1]{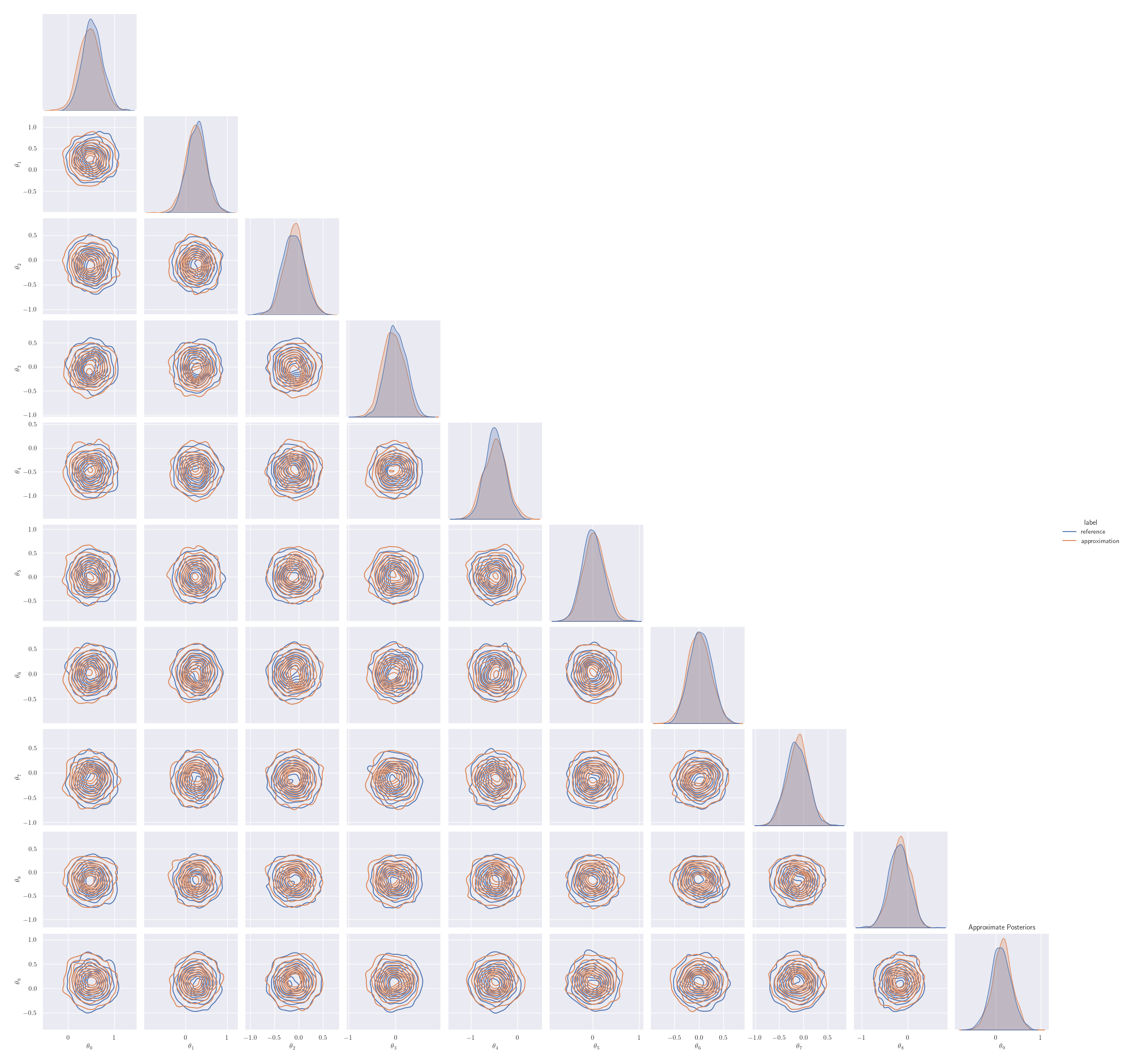}
\centering
\end{figure}

\subsection{Gaussian Mixture}
\begin{figure}[H]
\includegraphics[scale=0.4]{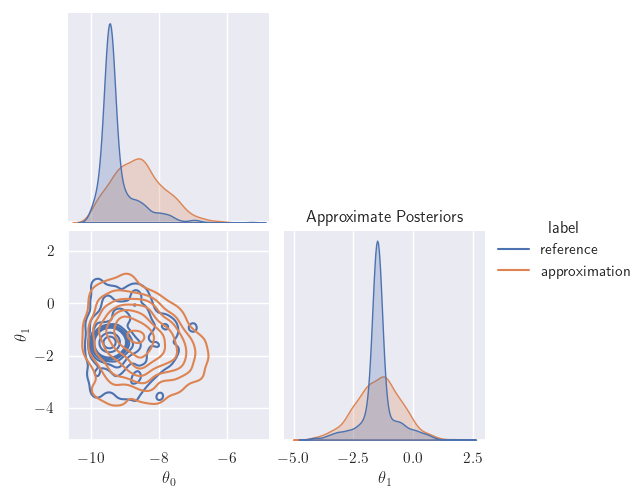}
\centering
\end{figure}

\subsection{Gaussian Linear Uniform}
\begin{figure}[H]
\includegraphics[scale=0.1]{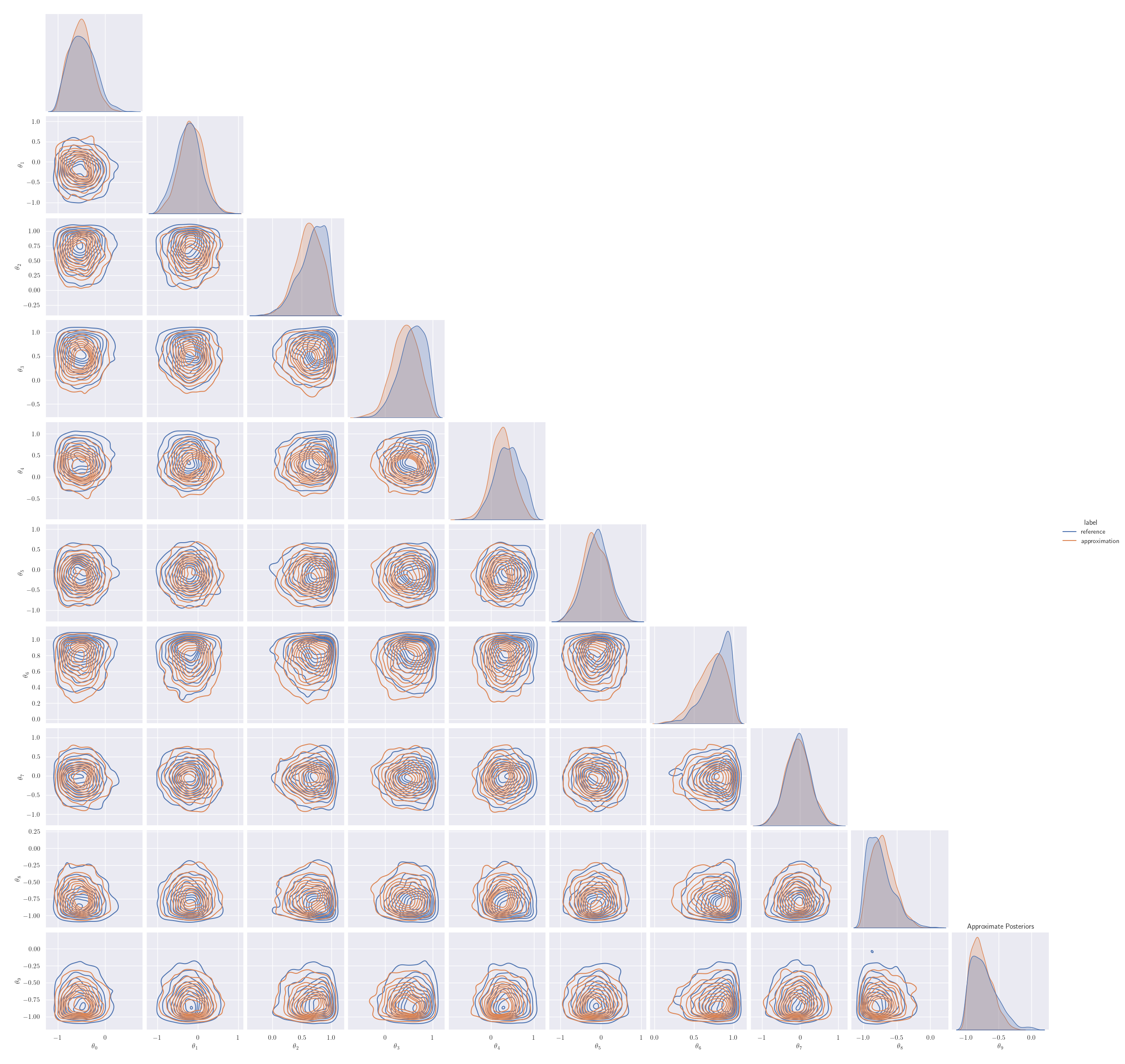}
\centering
\end{figure}

\subsection{Two Moons}
\begin{figure}[H]
\includegraphics[scale=0.4]{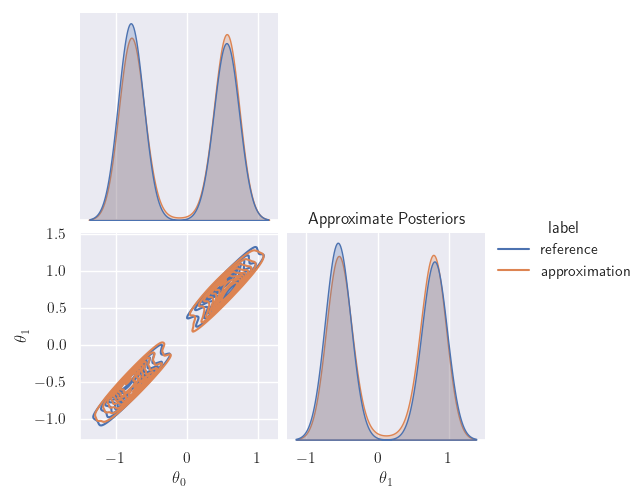}
\centering
\end{figure}

\subsection{SLCP}
\begin{figure}[H]
\includegraphics[scale=0.20]{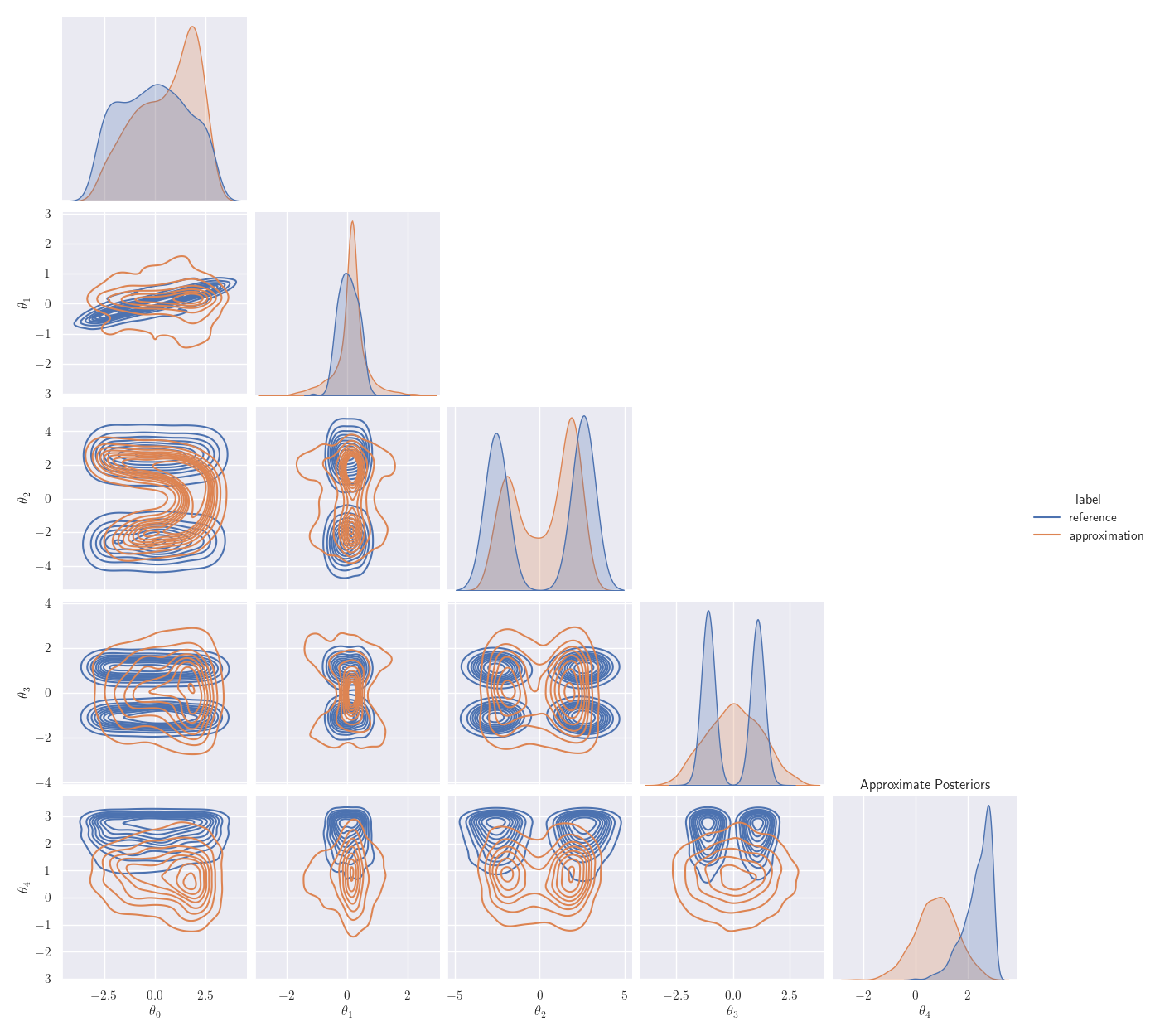}
\centering
\end{figure}


\subsection{Bernoulli GLM}
\begin{figure}[H]
\includegraphics[scale=0.1]{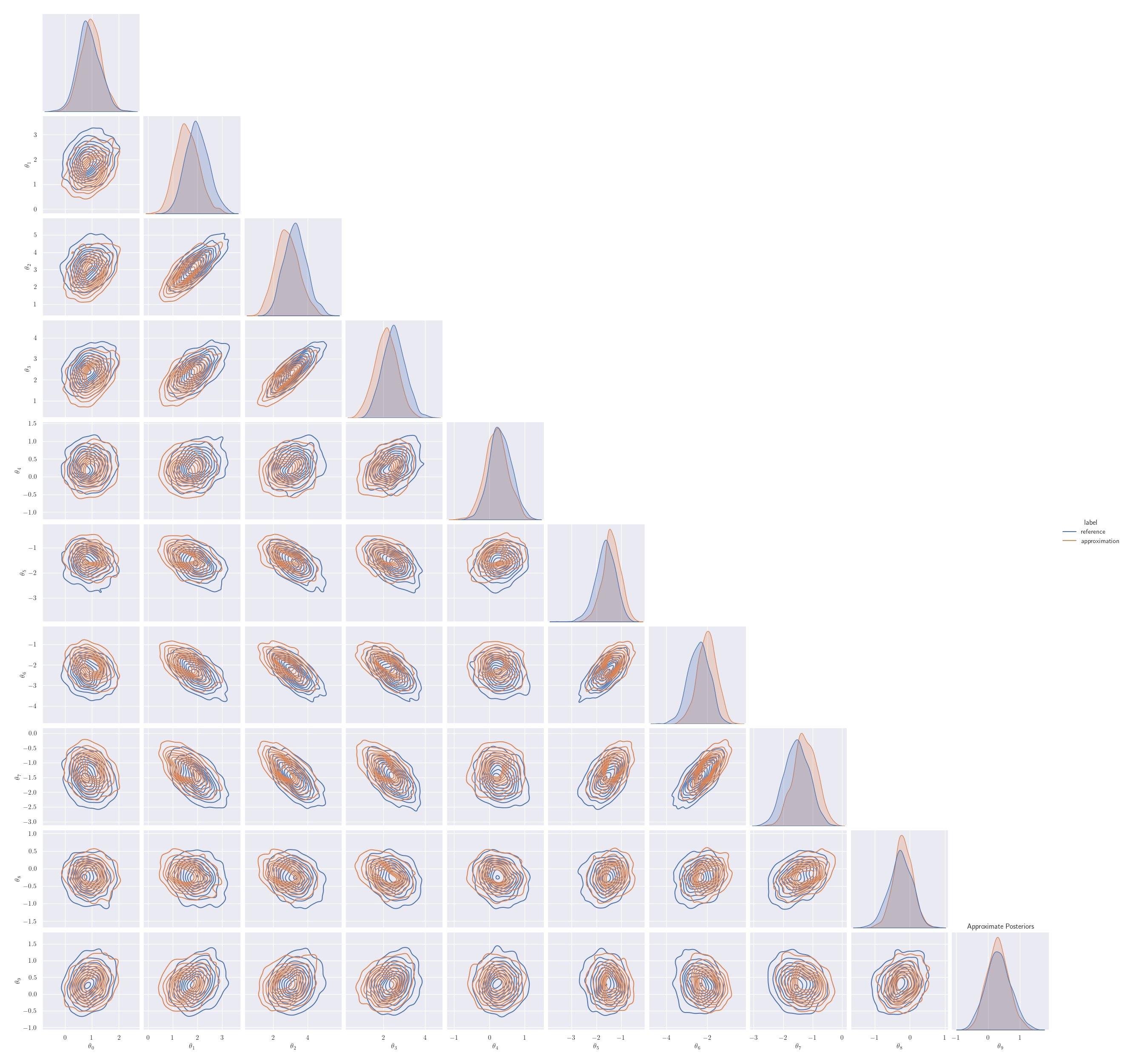}
\centering
\end{figure}

\subsection{ARCH}\label{section:arch_post}

For selected points from the training sets, we show the exact and approximate posteriors obtained by training according to the ELBO, IWBO, and FAVI objectives. The ground-truth parameter value is noted in red.

\begin{figure}[H]
    \centering
    \includegraphics[scale=.10]{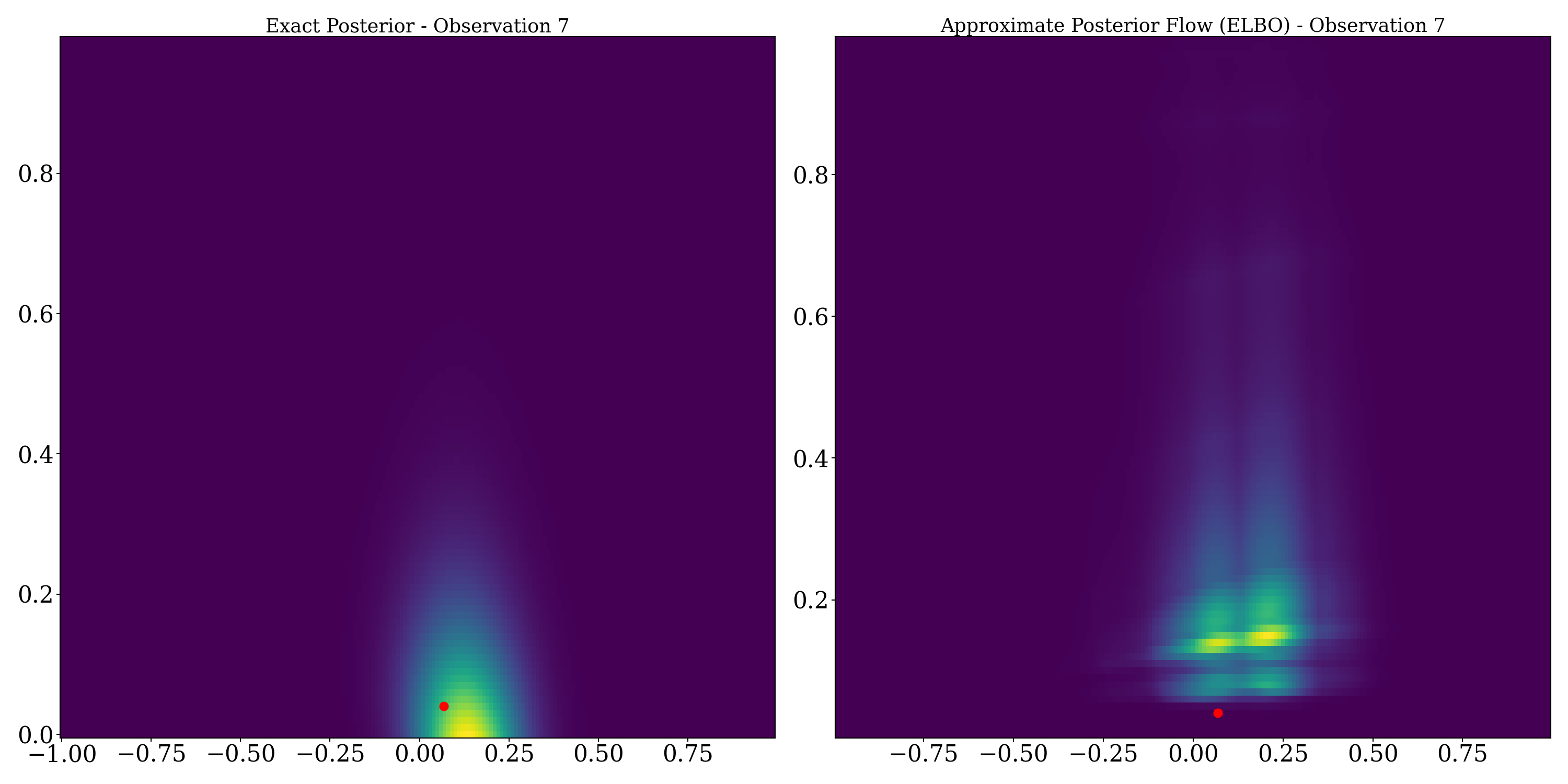}
\end{figure}
\begin{figure}[H]
    \centering
    \includegraphics[scale=.10]{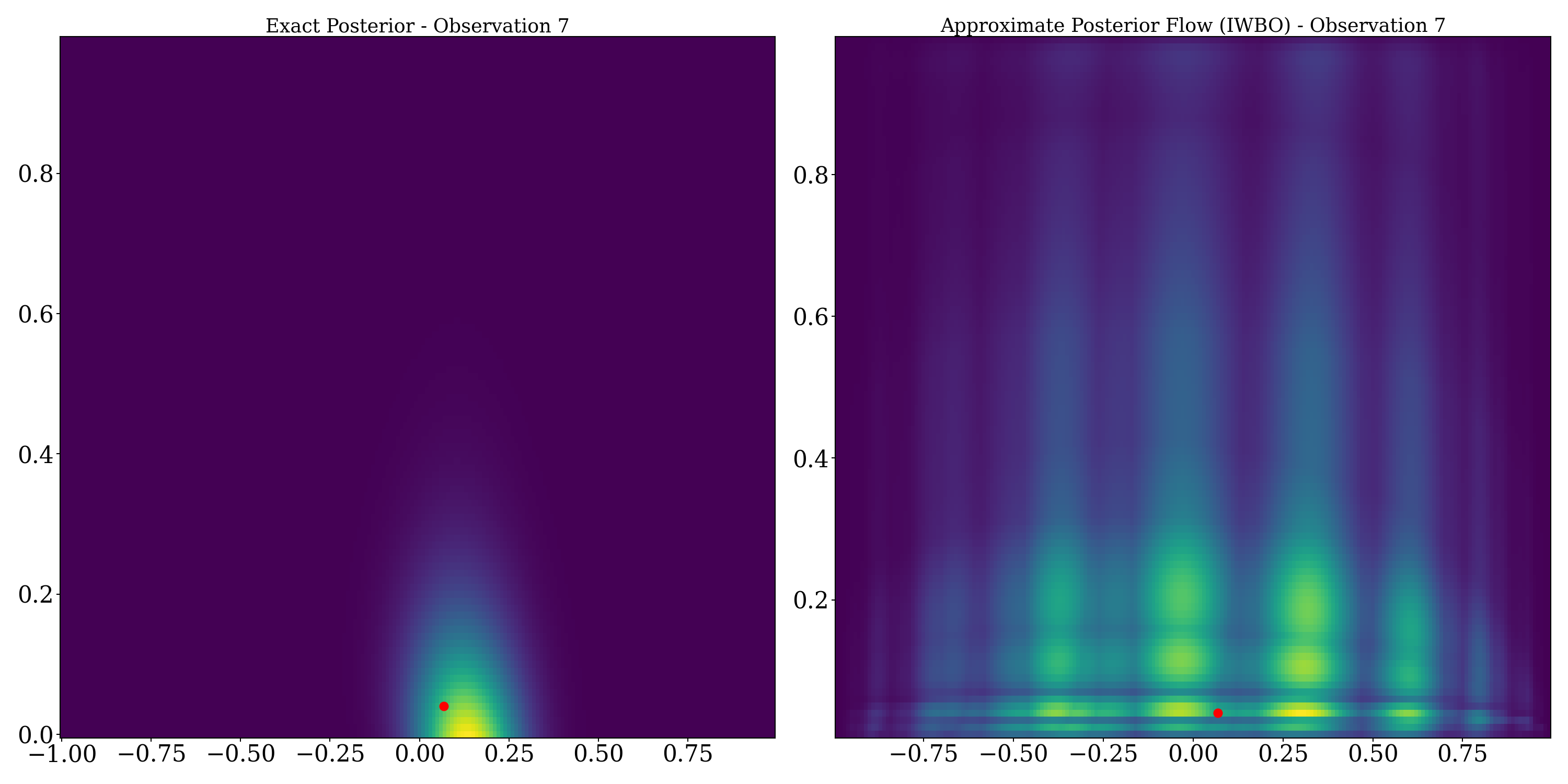}
\end{figure}
\begin{figure}[H]
    \centering
    \includegraphics[scale=.10]{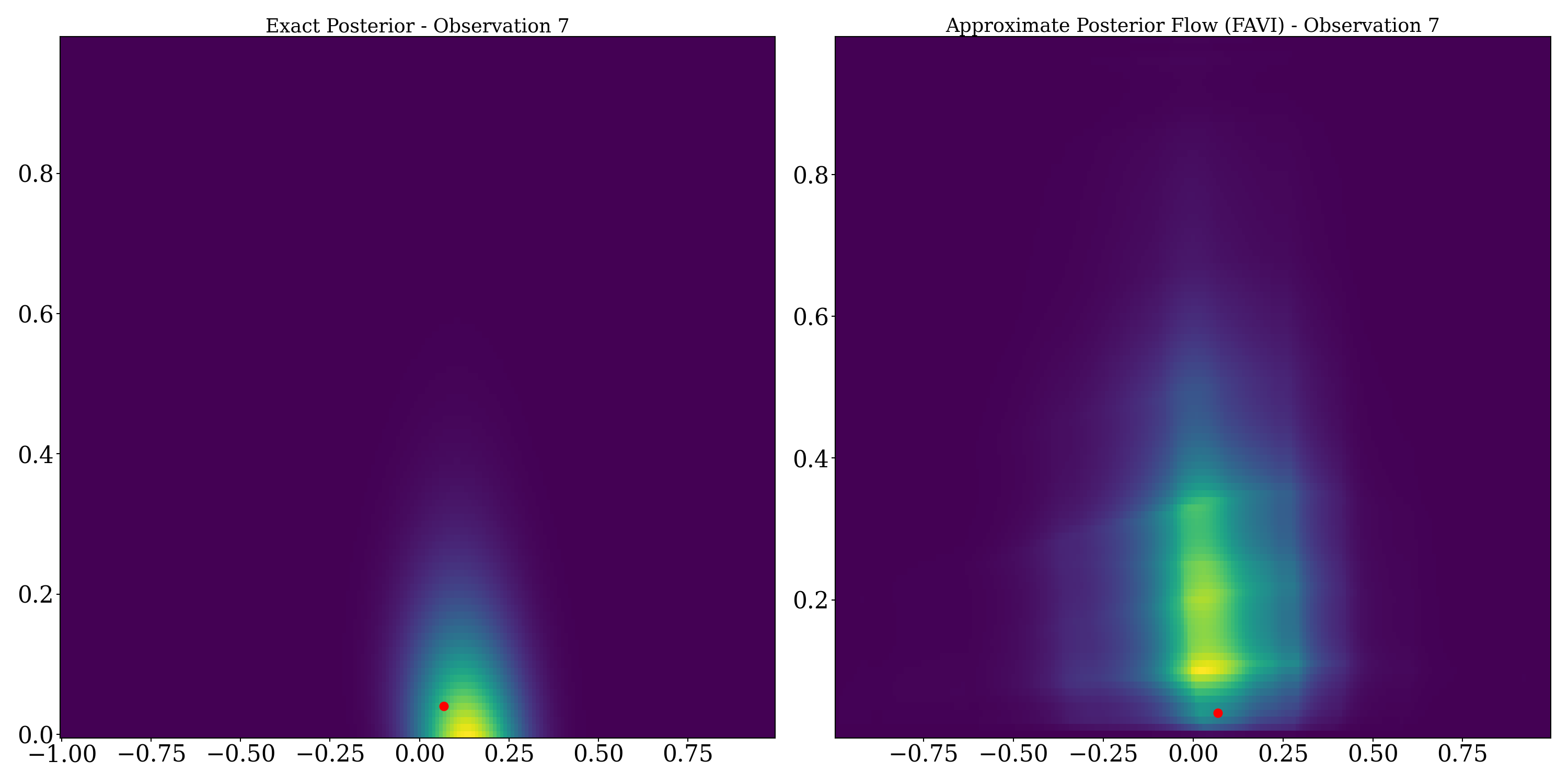}
\end{figure}
\begin{figure}[H]
    \centering
    \includegraphics[scale=.10]{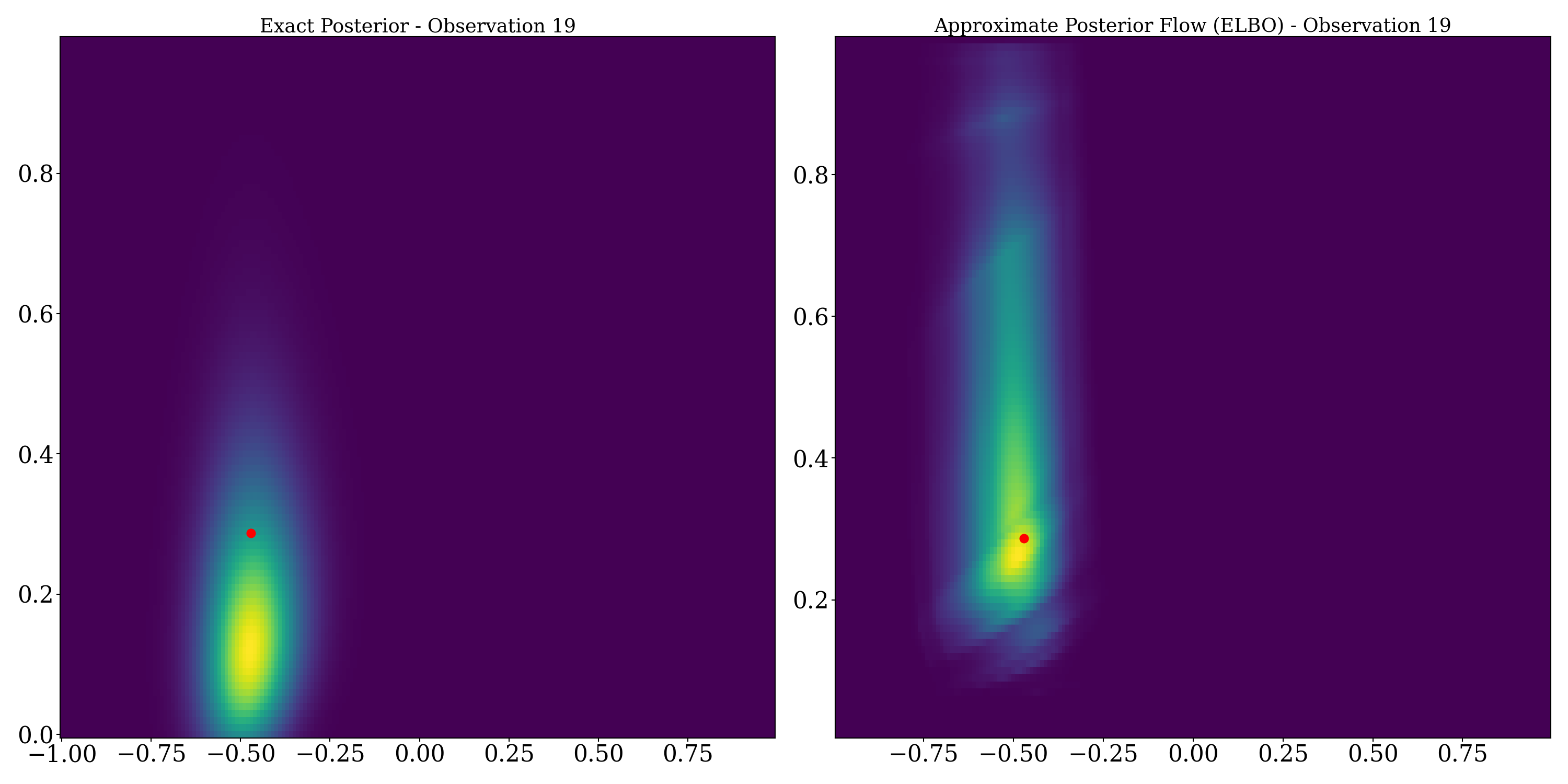}
\end{figure}
\begin{figure}[H]
    \centering
    \includegraphics[scale=.10]{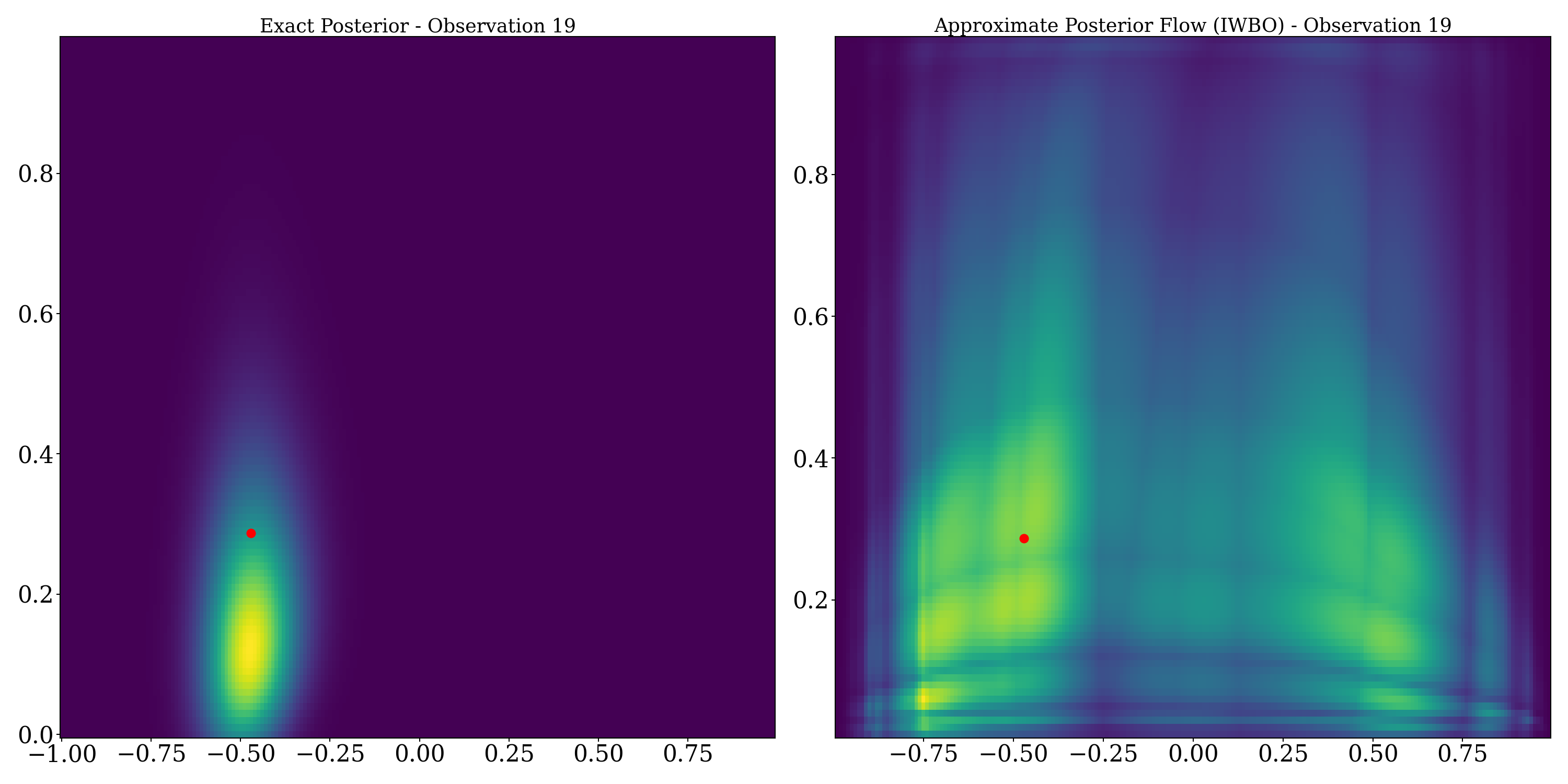}
\end{figure}
\begin{figure}[H]
    \centering
    \includegraphics[scale=.10]{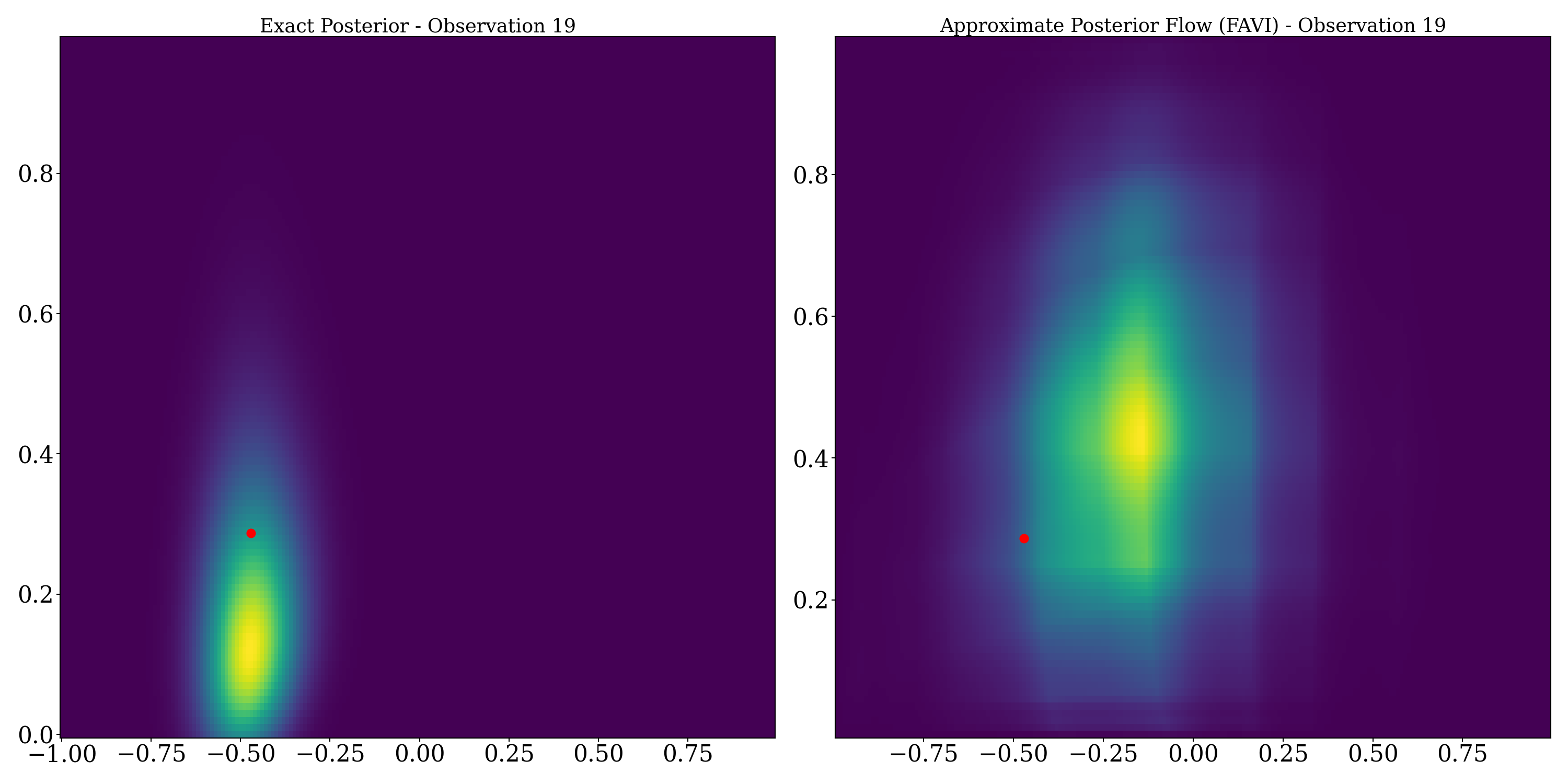}
\end{figure}



\begin{figure}[H]
\centering
  \includegraphics[scale=0.10]{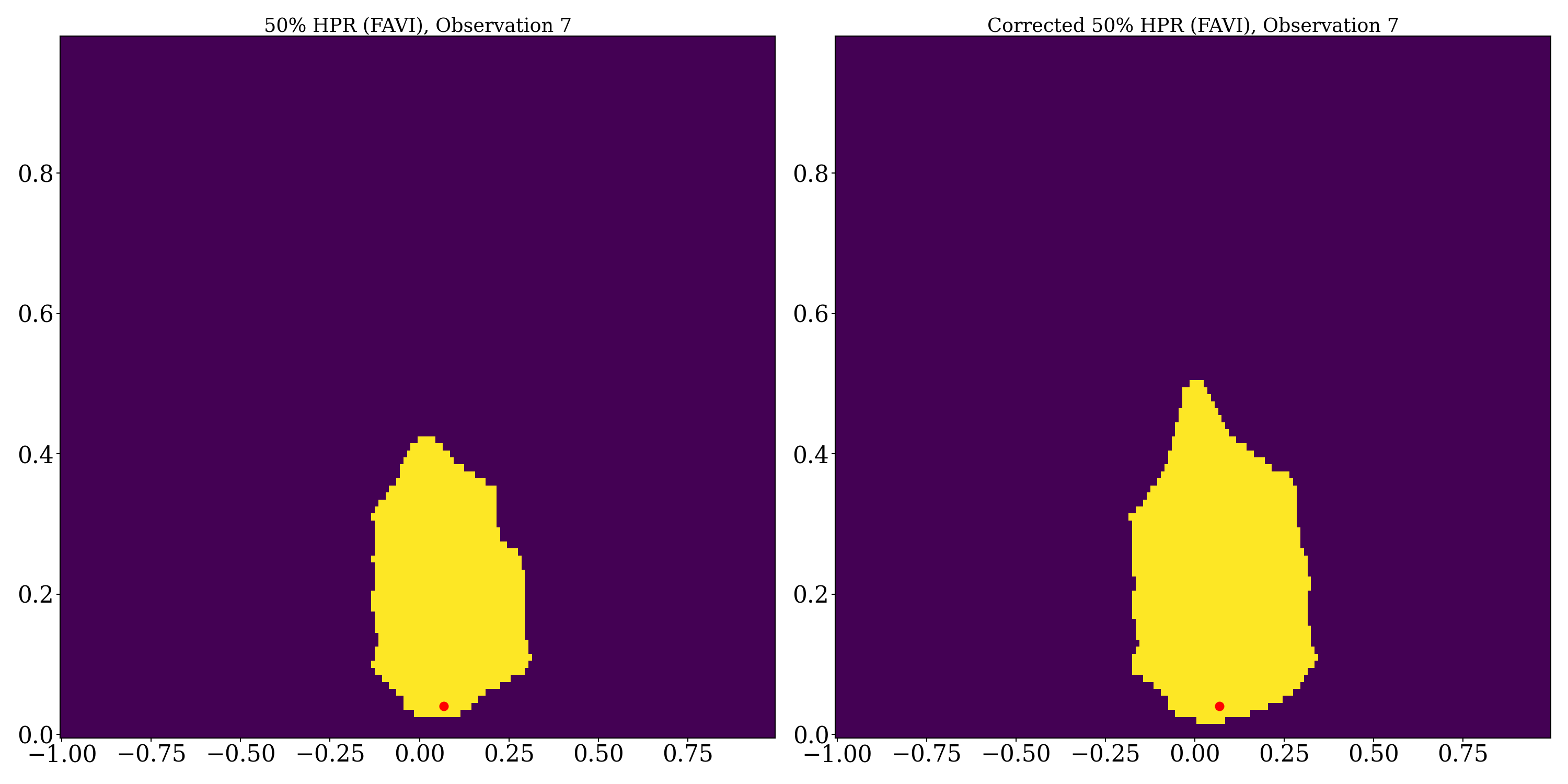}
  \caption{50\% prediction region, observation 7, before and after applying \textsc{CANVI}.}
  \label{fig:favi_shrink}
\end{figure}

\begin{figure}[H]
\centering
  \includegraphics[scale=0.10]{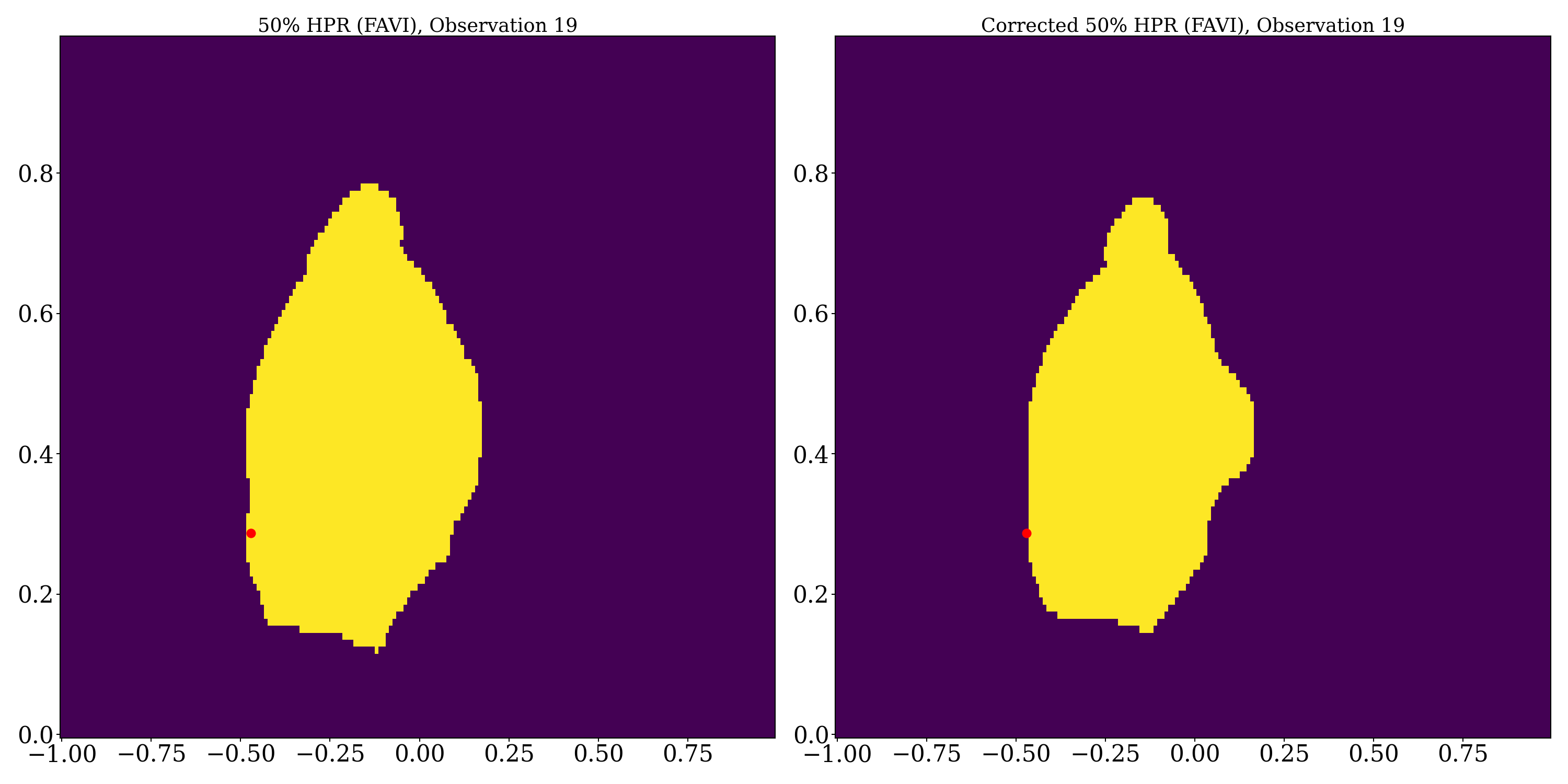}
\caption{50\% prediction region, observation 8, before and after applying \textsc{CANVI}.}
\label{fig:favi_grow}
\end{figure}

Depending on the miscalibration of the variational posterior on either per-objective or per-point basis, the conformalized $1-\alpha$ high-density region (HDR) either shrinks or expands, and can be visualized in two dimensions. \Cref{fig:favi_shrink}, \Cref{fig:favi_grow} show examples where the 50\% prediction regions obtained from the FAVI-learned variational posterior shrinks and grows after applying \textsc{CANVI}. 


\begin{figure}[H]
    \centering
    \includegraphics[scale=0.13]{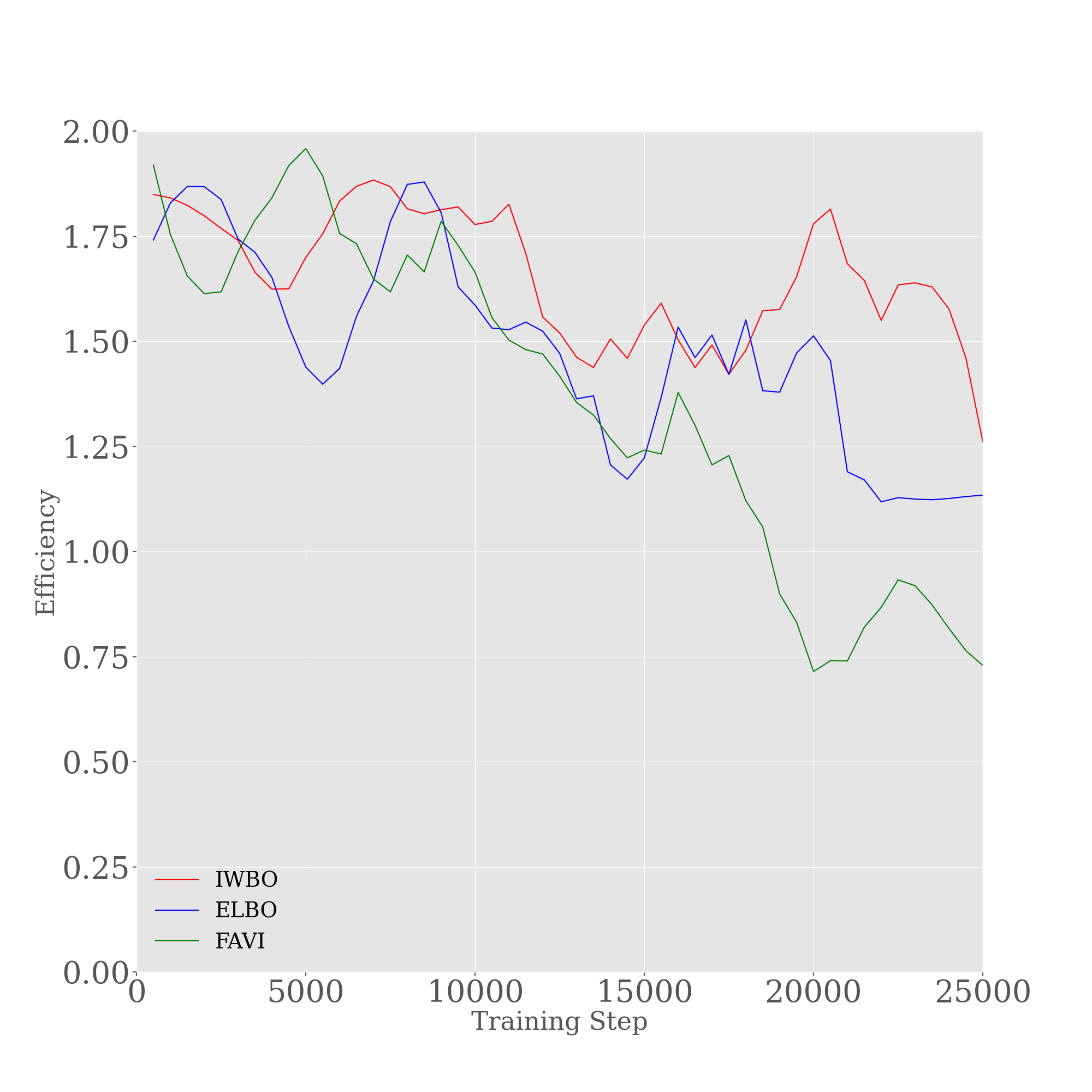}
    \caption{Efficiency estimates for variational posteriors trained by ELBO, IWBO, and FAVI across training iterations. }
    \label{fig:arch_eff}
\end{figure}

We visualize the efficiency of the iterates $q_\varphi$ across training iterations in Figure \ref{fig:arch_eff}. Recalling the form of Equation \ref{eqn:cp_mc_est}, it is unsurprising that FAVI tends to be more efficient, as it trains using simulated data (over which Equation \ref{eqn:cp_mc_est} is computed). To estimate Equation \ref{eqn:cp_mc_est}, at every 500 training steps, we simulate 20 $\theta,x$ pairs from the forward model. A larger number of Monte Carlo samples can be used but at an increased cost. As the resulting estimates are noisy, we smooth the resulting series with a Savitzky-Golay filter with a window length of 10 and third-degree polynomial order for better visualization.

\section{SED Experimental Details}\label{section:sed_setup}
\begin{figure}[H]
  \centering
  \includegraphics[scale=.20]{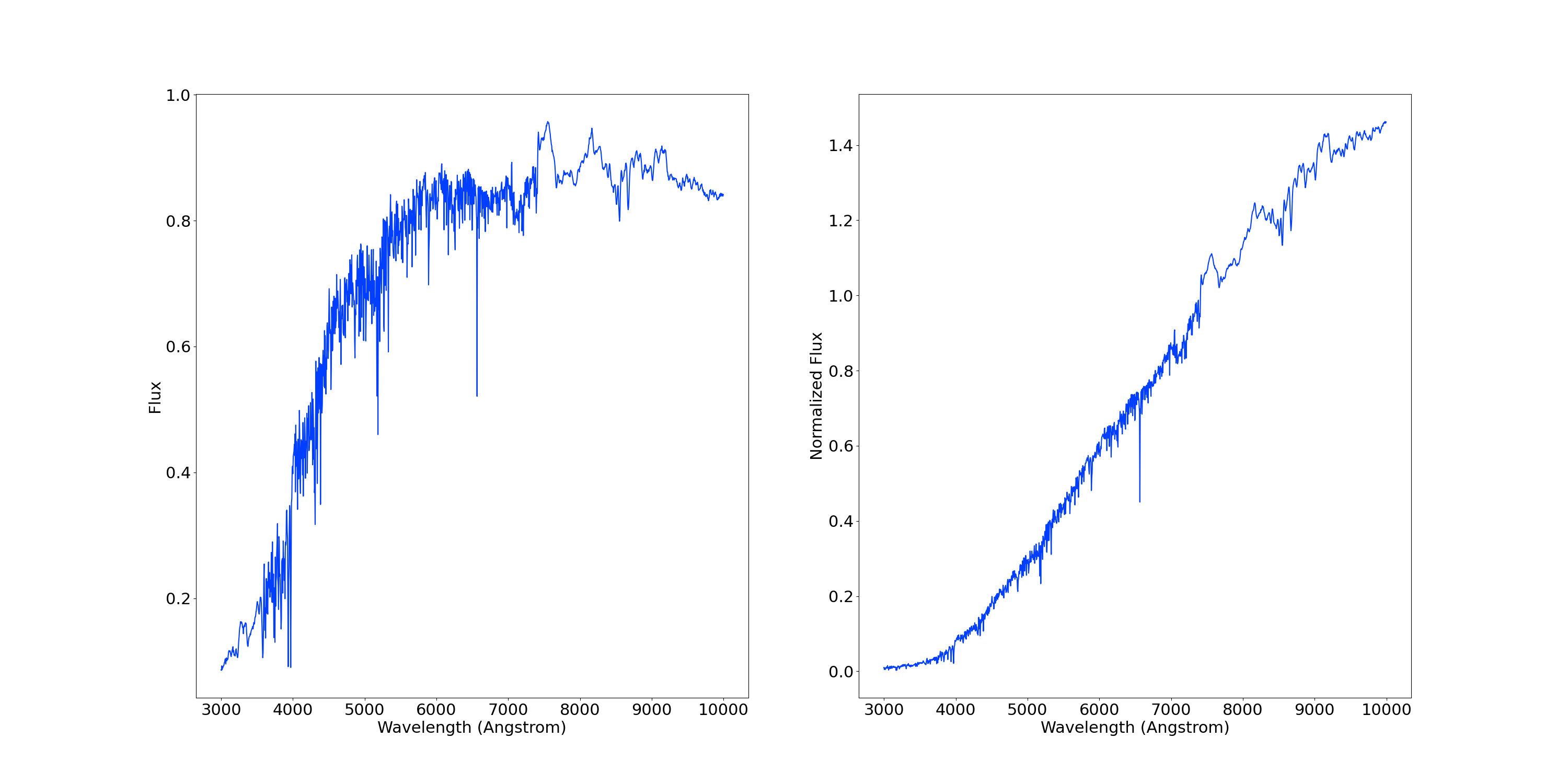}
  \caption{Two example draws from our neural network emulator of PROVABGS.}
  \label{fig:example_emulator_draws}
\end{figure}

The PROVABGS emulator (\Cref{subsection:seds}) was trained to minimize the MSE using normalized simulated PROVABGS outputs with fixed log stellar mass parameter \citep{hahn2023desi}. Training data were generated from PROVABGS with a fixed magnitude parameter ($\theta_0 = 10.5$), resampled onto a 5 Angstrom grid, and normalized to integrate to one. After training, forward passes through the emulator are significantly faster than the base simulator. We provide two simulated draws from our emulator in Figure \ref{fig:example_emulator_draws}. We use the recommended priors from \citep{hahn2023desi} on the remaining eleven parameters. As these are highly constrained (uniform priors, vastly different scales, and a 4-dimensional vector on the simplex), we similarly operate on an unconstrained, 10-dimensional space by invertible transformations. All three methods were trained for 10,000 steps using the Adam optimizer with learning rate 0.0001. $K=1000$ was used for the IWBO.

We let the output of our simulator be a mean parameter $\mu \in \mathbb{R}^{1400}$, and generate observed data independently binwise via $x_i \sim \mathcal{N}(\mu_i, |\mu_i|^2\sigma^2)$ for a fixed hyperparameter $\sigma=.1$ and all $i=1,\dots,1400$. We adopt this noise model for simplicity, as it imposes a fixed signal-to-noise ratio (SNR) of $1/\sigma = 10$  across all spectra and wavelength bins. 

\end{document}